\documentclass[prl,aps,onecolumn]{revtex4}

\usepackage{bm}
\usepackage{multirow}
\usepackage{ctable}
\usepackage{epstopdf}
\usepackage{placeins}
\usepackage[greek,dutch,english]{babel}
\usepackage{amsmath}                              
\usepackage{amssymb}                              
\usepackage[small,bf]{caption}                    
\usepackage{fancyhdr}                             
\usepackage{graphicx}                             
\usepackage{tabularx}                             
\usepackage{subfigure}                            
\usepackage{color}
\usepackage{natbib}
\usepackage{makeidx}

\usepackage{chngcntr}
\usepackage{bbm}                                  
\usepackage{textcomp}

\newcommand{\ket}[1]{\ensuremath{|#1\rangle}}
\newcommand{\kxy}[1]{\ensuremath{|\text{#1}\rangle}}
\newcommand{\bra}[1]{\ensuremath{\langle#1|}}

\newcommand{\ev}[1]{\ensuremath{\langle#1\rangle}}
\def\be{\begin{equation}}
\def\ee{\end{equation}}

\newcommand{\beginsupplement}{%
        \setcounter{table}{0}
        \renewcommand{\thetable}{\arabic{table}}%
        \setcounter{figure}{0}
        \renewcommand{\thefigure}{\arabic{figure}}%
	}

\begin{document}

\title{Repeated quantum error correction on a continuously encoded qubit
by real-time feedback}

\author{J. Cramer$^{1,2}$, N. Kalb$^{1,2}$, M. A. Rol$^{1,2}$, B. Hensen$^{1,2}$, M. S. Blok$^{1,2}$, M. Markham$^3$, D. J. Twitchen$^3$,  R. Hanson$^{1,2}$, T.~H.~Taminiau$^{1,2*}$}
\affiliation{
$^1$QuTech, Delft University of Technology, PO Box 5046, 2600 GA Delft, The Netherlands.\\
$^2$Kavli Institute of Nanoscience, Delft University of Technology, PO Box 5046, 2600 GA Delft, The Netherlands.\\
$^3$Element Six Innovation, Fermi Avenue, Harwell Oxford, Didcot, Oxfordshire OX11 0QR, United Kingdom.\\
$^*$Correspondence to: T.H.Taminiau@TUDelft.nl}

\begin{abstract}
Reliable quantum information processing in the face of errors is a major fundamental and technological challenge. Quantum error correction protects quantum states by encoding a logical quantum bit (qubit) in multiple physical qubits. To be compatible with universal fault-tolerant computations, it is essential that the states remain encoded at all times and that errors are actively corrected. Here we demonstrate such active error correction on a continuously protected qubit using a diamond quantum processor. We encode a logical qubit in three long-lived nuclear spins, repeatedly detect phase errors by non-destructive measurements using an ancilla electron spin, and apply corrections on the encoded state by real-time feedback. The actively error-corrected qubit is robust against errors and multiple rounds of error correction prevent errors from accumulating. Moreover, by correcting correlated phase errors naturally induced by the environment, we demonstrate that encoded quantum superposition states are preserved beyond the dephasing time of the best physical qubit used in the encoding. These results establish a powerful platform for the fundamental investigation of error correction under different types of noise and mark an important step towards fault-tolerant quantum information processing.\end{abstract}

\maketitle

Large-scale quantum information processing requires the correction of errors during computations. In quantum error correction a logical quantum bit (qubit) is encoded in a subspace of multiple physical qubits so that errors can be actively corrected without affecting the encoded information. A promising way to correct errors in encoded quantum states is to perform feedback based on multi-qubit measurements known as stabilizer measurements [1-3] (see Fig. 1a). These measurements are performed non-destructively using extra qubits (ancillas) and are frequently repeated to detect errors before they accumulate. The measurement outcomes are then processed in classical logic that identifies the error syndrome, and, in order to enable universal computations [1], active feedback is applied to the encoded system to correct errors where needed. The key experimental challenge is to perform such complete error-correction cycles including non-destructive stabilizer measurements and real-time feedback well within the coherence time. 


Quantum-error-correction protocols have been explored across a range of platforms [4-14]. Pioneering experiments bypassed stabilizer measurements by reversing the encoding to correct errors, thus leaving the quantum state unprotected [5-11]. Recent breakthroughs have enabled the use of stabilizer measurements to passively track errors in quantum states and retrieve stored information afterwards through post processing [12-15]. Here we realize complete rounds of active quantum error correction on a continuously encoded qubit by exploiting newly-developed stabilizer measurements based on an electron spin ancilla with high-fidelity non-demolition readout, encoding in long-lived nuclear spins and real-time correction of errors through fast classical logic. 


The three-qubit code considered here corrects a single phase error on any one of the physical qubits. The logical qubit is encoded as $\ket{\psi}_L=\alpha\ket{0}_L+\beta\ket{1}_L$ with $\ket{0}_L=(\kxy{+X}_1\kxy{+X}_2\kxy{+X}_3+\kxy{-X}_1\kxy{-X}_2\kxy{-X}_3)/\sqrt{2}$, $\ket{1}_L=(\kxy{+X}_1\kxy{+X}_2\kxy{+X}_3-\kxy{-X}_1\kxy{-X}_2\kxy{-X}_3)/\sqrt{2}$ and $\ket{\pm\text{X}}=(\ket{0}\pm\ket{1})/\sqrt{2}$. The logical qubit operators are $X_L=X_1I_2I_3$, $Y_L=Y_1Z_2Z_3$ and $Z_L=Z_1Z_2Z_3$ (or their permutations). Errors are detected by measuring the stabilizer generators $X_1X_2I_3$ and $I_1X_2X_3$ via an ancilla. For an uncorrupted state both measurements yield outcome +1 (no error), but for a phase error (a $Z$ operation) on just one of the qubits the two measurements give a unique syndrome of -1 outcomes that identifies the error.
\newline
\textbf{RESULTS}

\textbf{Stabilizer measurements and real-time feedback.} Our qubits are three $^{13}$C nuclear spins ($I=1/2$, $1.1\%$ abundance) surrounding a single nitrogen-vacancy (NV) centre in diamond, whose electronic spin we use as ancilla ($S=1$; $\ket{0}_a:m_s=0$ and $\ket{1}_a:m_s=-1$) (Fig. 1b). At 4 K, the ancilla combines fast control [16], optical single-shot readout [17] and long coherence times [18] ($> 25$ ms, Methods). We use relatively remote nuclear qubits (coupling to the ancilla 20-50 kHz) that are robust against optical excitation of the ancilla and design decoherence-protected gates to control them [9,19] (Methods). All three qubits show long dephasing times $T_2^*$ (Fig. 1c) with the dominant natural errors being phase errors.


The key challenge for implementing stabilizer measurements in this system is that the ancilla-qubit interaction is always present: imperfect knowledge of the ancilla state during or after readout dephases the qubits [20-22]. To minimize this dephasing, we implement quantum non-demolition measurements of the ancilla by resonant optical excitation of $\ket{0}_a$ and stopping the excitation within 2 $\mu$s upon photon detection (outcome $\ket{0}_a$) to minimize spin flips in the optically excited state [23] (Methods). The resulting readout fidelities are $F_0=0.890(4)$ for $\ket{0}_a$ and $F_1=0.988(2)$ for $\ket{1}_a$ (average: $F=0.939(2)$). Crucially, the post-measurement fidelity after correctly assigning $\ket{0}_a$ is 0.992, demonstrating the desired non-demolition character.


To benchmark the stabilizer measurements and real-time feedback, we deterministically entangle two qubits by projecting into a Bell state, i.e. a simultaneous eigenstate of $XX$ and $ZZ$ [21,24,25]. First, the qubits are initialized in $\ket{00}$, an eigenstate of $ZZ$, with fidelity 0.910(6). Then a $XX$ stabilizer measurement projects the qubits onto one of two Bell states (Fig. 1d). We interpret the -1 outcome as an error in the desired state and correct it through feedback before performing two-qubit tomography. The deterministically generated entangled state, with fidelity $F=0.824(7)$ (Fig. 1e), demonstrates the non-destructive nature of the measurement; coherence within the subspaces is maintained throughout the measurement and feedback cycle. The complete cycle can be repeated up to 6 times within the shortest qubit $T_2^*$.

\textbf{Active quantum error correction on a logical qubit.} We now turn to quantum error correction by stabilizer measurements. The logical qubit is encoded by taking an arbitrary state $\ket{\psi}_a=\alpha\ket{0}_a+\beta\ket{1}_a$ prepared on the ancilla to the three-qubit state $\ket{\psi}_L=\alpha\ket{0}_L+\beta\ket{1}_L$ (Fig. 2a). We characterize the encoding by preparing six basis states $\ket{0}_L$, $\ket{1}_L$, $\ket{\pm \text{X}}_L=(\ket{0}_L\pm\ket{1}_L)/\sqrt{2}$ and $\ket{\pm \text{Y}}_L=(\ket{0}_L\pm i\ket{1}_L)/\sqrt{2}$ and performing three-qubit state tomography. The fidelities with the ideal states confirm successful encoding and genuine three-qubit entanglement (Fig. 2b).


We first investigate the recovery of arbitrary logical qubit states from phase errors. To emulate a general process causing dephasing, uncorrelated incoherent errors are applied with variable probability $p_e$ to each physical qubit simultaneously (Fig. 3a); for each qubit the error process is $E(\rho)=(1-p_e)I\rho I+p_eZ\rho Z$, with $\rho$ the single-qubit density matrix. By controllably applying such errors we characterize the effectiveness of the error correction for any process causing uncorrelated errors with equal probability to the qubits. We then measure the stabilizers $X_1X_2I_3$ and $I_1X_2X_3$, identify potential errors and correct them through feedback. The probabilities to obtain the four different error syndromes (inset in Fig. 3b) show the expected symmetry around $p_e=0.5$ and match the theoretical prediction based on the errors present in the initial states (Fig. 2b) and the average ancilla readout fidelity.


The protection of the logical qubit is characterized by the process fidelity with the identity (Fig. 3b) (Methods). We quantitatively analyse the results by fitting to $wF_{QEC}+(1-w)F_{linear}$, where $F_{QEC}(p_e)$ and $F_{linear}(p_e)$ are the theoretical curves with and without error correction ($w=1$ indicates ideal robustness against applied single-qubit errors). When no error correction is applied we observe the expected linear dependence on the error probability: $w\approx 0$. In contrast, with quantum error correction $w$ is $0.81(3)$, and a non-linear curve shape that is characteristic for robustness against single-qubit errors is obtained. This result demonstrates that the entropy associated to the applied errors is successfully removed from the system.


Comparisons to an unencoded qubit and the logical qubit without error correction reveal that adding quantum error correction on top of a computation does not yet provide a net improvement (Fig. 3b), due to additional errors introduced by the initialization, encoding and stabilizer measurements (total of 13 two-qubit gates, 488 ancilla refocusing pulses and 6 ancilla readouts/resets). To isolate the errors due to the stabilizer measurements, we compare the error-corrected logical qubit to the logical qubit left idle. We further optimize the error correction, by assigning the ancilla state with the best readout fidelity ($\ket{1}_a$, $F_1=0.988(2)$) to the most likely error syndrome (+1, +1 - no error, inset Fig. 3b), instead of averaging over all assignments as in Fig. 3b. With this improvement, error correction outperforms idling for a range of $p_e$ (Fig. 3c); once the logical qubit is encoded, quantum error correction can be beneficial.


\textbf{Multiple rounds of active error correction.} Because a complete round of error correction (2.99 ms) fits well within the dephasing time of the physical qubits, we can concatenate multiple rounds to improve the coherence of continuously encoded quantum superpositions by preventing the accumulation of errors (Fig. 4a). Three new elements are introduced. First, the total error probability $p_e$ is distributed over $n$ rounds, so that the error probability per round is $p_n=(1-\sqrt[n]{1-2p_e})/2$ (Methods). This error model corresponds to errors occurring incoherently, for example with a constant rate in time. Second, to investigate dephasing we consider only the protection of the two states $\ket{\pm \text{X}}_L=\ket{\pm \text{X},\pm \text{X},\pm \text{X}}$ (i.e. a classical bit stored in the phase of a quantum superposition). Third, we exploit the intrinsic robustness of the logical qubit to single $Z$ errors by redefining $X_L=(X_1I_2I_3+I_1X_2I_3+I_1I_2X_3-X_1X_2X_3)/2$ which is equivalent to performing a round of error correction by majority voting at the end of the experiment [13,14].


For a single round of error correction (majority vote only) the average fidelity is higher than for an unencoded qubit for any $p_e$ (Fig. 4b); adding more (identical) qubits is always beneficial in the repetition code. For $p_e=0$, additional rounds of quantum error correction can only introduce errors, reducing the fidelity (Fig. 4b). For larger $p_e$, however, multiple rounds prevent errors from accumulating by dividing the error process in parts that are more likely to contain only single errors, which are corrected. In addition, unlike error detection with post-processing [13,14], active correction between rounds keeps the probability to obtain +1 (no error) high (inset Fig. 4b) and thus maintains the advantage of assigning the highest-fidelity ancilla readout to that outcome. As a result, for  $p_e>0.3$, multiple rounds outperform a single round of error correction. 


\textbf{Correcting natural dephasing.} Finally, as an example of suppressing errors naturally present in the environment, we let the qubits evolve naturally instead of applying errors (Fig. 4c). The error probabilities are now different for each qubit because their intrinsic $T_2^*$ differ due to their local environments (Fig. 1c). In addition, the errors arise from quasistatic detunings due to the fluctuating $^{13}$C spin bath, so that they are correlated in time and can evolve coherently. Like most environmental errors, such errors might also be suppressed by other methods than quantum error correction, for example by polarizing the spin environment [26,27], by refocusing pulses [28] or by isotopic purification [28-31].


The fidelity for the logical qubit with majority voting again starts above the best unencoded qubit, but drops below it for larger evolution times (Fig. 4d). Because the error probabilities vary between qubits, an error detected on the best qubit becomes more likely to actually correspond to errors on both other qubits and the wrong correction is made. An additional round of quantum error correction in the middle of the evolution time now not only prevents errors from accumulating by intermediately correcting them, but also interrupts any coherent build-up by projecting the errors, thus suppressing them (Fig. 4d). Due to this combination, the logical qubit with error correction yields the highest average fidelity for total evolution times between 5 and 19 ms (Fig. 4d). This result demonstrates an actively error-corrected logical qubit with an improved dephasing time over the best qubit used in the encoding.
\newline
\textbf{DISCUSSION}

The presented non-destructive measurements and real-time feedback on encoded quantum states are the key primitives for universal computations on logical qubits and for error-correcting codes that correct both phase and bit-flip errors. To reach scalability thresholds, readout and gate fidelities can be further improved, for example through optical cavities [32], the implantation [33] or selective growth of defects and isotopes [28,29], and optimal control [33]. In a wider perspective, our results can be combined with recently demonstrated entanglement between distant NV centres [34,35] to form quantum networks with error-corrected nodes for entanglement purification, quantum communication and networked quantum computation [36]. Therefore, these results establish a promising platform to experimentally investigate protocols for fault-tolerant quantum information processing under different types of noise and error correlations in diverse settings.

\section{METHODS}

\textbf{Sample and Setup.} We use a naturally occurring Nitrogen-Vacancy (NV) in high-purity type IIa chemical-vapour-deposition (CVD) grown diamond with a $1.1\%$ natural abundance of $^{13}$C and a $<111>$ crystal orientation (Element Six). To enhance the collection efficiency a solid-immersion lens was fabricated on top of the NV centre [17,37] (Fig. 1b) and a single-layer aluminium-oxide anti-reflection coating was deposited [34,38]. The sample temperature is $T\approx4.2$  K and a magnetic field of 403.553(3) G is applied along the NV symmetry axis.


The ancilla NV electron spin is characterized by a Rabi frequency of 4.3 MHz, a dephasing time $T_2^*$=4.6(2)  $\mu$s, a Hahn echo time $T_2=1.03(3)$  ms and a longitudinal relaxation time of 0.43(6) s (due to microwave noise and laser background). The coherence time of the ancilla under dynamical decoupling exceeds 25 ms and does not limit the experiments (Supplementary Fig. 1). We initialize and readout the ancilla through resonant excitation of the zero-phonon transitions of the NV centre (Supplementary Fig. 2). Prior to every experiment the $^{14}$N nuclear spin is initialized by measurement with a fidelity of $F_N$ = 0.94(3) in $m_I=-1$ [17]. No external electric fields are applied: the gates in Fig. 1b are grounded.


\textbf{Nuclear spin qubit control.} The hyperfine interactions for the three nuclear spins are estimated by dynamical decoupling spectroscopy [9] (Supplementary Table 1). Building on previous gate designs [9], nuclear gates are realized by applying sequences of $\pi$-pulses on the electron spin of the form $(\tau - \pi - 2\tau -  \pi - \tau)^{N/2}$. The number of pulses $N$ sets the rotation angle. The inter-pulse delay $2\tau$ determines which qubit is controlled and whether the rotation is conditional on the ancilla state. In contrast to previous work [9] we allow the gates to be detuned, providing greater flexibility to optimize $\tau$ and $N$ for gate selectivity and minimal discretization errors. The gate parameters are listed in Supplementary Table 1 and 2.


The nuclear spins are initialized by swapping with the ancilla electron spin (Supplementary Fig. 3) and are read out by mapping the required correlation to the ancilla before reading it out (Supplementary Fig. 4). To obtain best estimates for the actual states, the results are corrected for the fidelity of the gates used in the final readout (tomography) (details in Supplementary Note 3). Uncorrected data is shown in Supplementary Fig. 11.


\textbf{Feedback.} 
Real-time feedback is implemented through a programmable microprocessor (ADwin Pro II) that controls the experimental sequence (Supplementary Fig. 5). We exploit feedback in four different ways. First, detected phase errors are corrected directly after the stabilizer measurements. Note that analysing errors over multiple rounds [14] would additionally enable real-time correction of ancilla readout errors, but that this is not implemented here. Second, dependent on the ancilla measurement outcome the qubits pick up a deterministic phase shift due to the hyperfine interaction, which is corrected in the same way. Third, for an odd number of +1 outcomes the operations in the stabilizer measurements imprint a bit flip on the logical qubit, which we correct by transforming the logical qubit basis in real time. Fourth, to start each measurement sequence with the ancilla in $\ket{0}_a$ it is flipped back to $\ket{0}_a$ when the previous measurement returned $\ket{1}_a$. 

Importantly, we perform real-time feedback either by adapting the qubit bases for all subsequent gates and measurements (for correcting $Z$ errors and for the logical qubit) or by absorbing the feedback operations into the next gate acting on the same qubit (for the ancilla). Therefore the physical control sequence is directly adapted based on the measurement outcomes without introducing unnecessary gate operations that would cause additional errors. In the circuit diagrams we sometimes display the gates for the feedback separately for clarity.


\textbf{Quantum error correction analysis.} The process fidelity with the identity is given by $F_p=(F_0+F_1+F_{+X}+F_{-X}+F_{+Y}+F_{-Y}-2)/4$, with $F_{\alpha} = \bra{\alpha}\rho_{\alpha}\ket{\alpha}$ the six fidelities of the final states $\rho_{\alpha}$ with the ideal states $\ket{\alpha}_L$. The results of Fig. 3 are analysed by fitting to $wF_{QEC}(p_e)+(1-w)F_{linear}(p_e)$, with $F_{QEC}(p_e)=O+A(1-3p_e^2+2p_e^3)$ and $F_{linear}(p_e) =O+A(1-p_e)$. $A$ and $O$ account for the experimental fidelities (Supplementary Note 1). 

The state fidelities for multiple rounds of error correction and incoherent errors (Fig. 4b) are fitted to the same equation using $F_{QEC}(p_e)=\frac{1}{2}(1+A(1-6p_n^2+4p_n^3)^n)$, with $n$ the number of rounds, $p_n=\frac{1}{2}\sqrt[n]{1-2p_e}$ the error per round, and $F_{linear}(p_e)=\frac{1}{2}(1+A(1-2p_e))$. The error per round $p_n$ is obtained as follows. An error process with total error probability ($p_e$) reduces  the expectation value by a factor of $(1-2p_e)$. For incoherent errors, a process can be divided in $n$ equal rounds using $(1-2p_e)=(1-2p_n)^n$, which results in $p_n = (1-\sqrt[n]{1-2p_e})/2$ (for $p_e \leq 0.5$). In Fig. 3c and Fig. 4b, $A$ depends on the error-probability $p_e$, because we optimize the effective readout fidelity by associating the most likely error syndrome to the best ancilla readout (Supplementary Note 1). See Supplementary Information for further details on all theoretical analysis, including the error syndrome probabilities and numerical simulations of Fig. 4d.

\section{References}

\noindent [1] B. M. Terhal, Quantum error correction for quantum memories. \textit{Rev. Mod. Phys. }\textbf{87}, 307 (2015).

\noindent [2] S. B. Bravyi and A. Yu. Kitaev, Quantum codes on a lattice with boundary. arXiv: 9811.052 (1998).

\noindent [3] R. Raussendorf, J. Harrington, Fault-tolerant quantum computation with high threshold in two dimensions. \textit{Phys. Rev. Lett.} \textbf{98}, 190504 (2007).

\noindent [4] D. Nigg \textit{et al.}, Quantum computations on a topologically encoded qubit. \textit{Science} \textbf{345}, 302-305 (2014).

\noindent [5] E. Knill, R. Laflamme, R. Martinez, C. Negrevergne, Benchmarking quantum computers: the five-qubit error correcting code. \textit{Phys. Rev. Lett.} \textbf{86}, 5811-5814 (2001).

\noindent [6] J. Chiaverini \textit{et al.}, Realization of quantum error correction. \textit{Nature} \textbf{432}, 602-605 (2004).

\noindent [7] P. Schindler \textit{et al.}, Experimental repetitive quantum error correction. \textit{Science} \textbf{332}, 1059-1061 (2011).

\noindent [8] M. D. Reed \textit{et al.}, Realization of three-qubit quantum error correction with superconducting circuits. \textit{Nature} \textbf{482}, 382-385 (2012).

\noindent [9] T. H. Taminiau, J. Cramer, T. van der Sar, V. V. Dobrovitski, R. Hanson, Universal control and error correction in multi-qubit spin registers in diamond. \textit{Nature Nanotech.} \textbf{9}, 171-176 (2014).

\noindent [10] G. Waldherr \textit{et al.}, Quantum error correction in a solid-state hybrid spin register. \textit{Nature} \textbf{506}, 204-207 (2014).

\noindent [11] B. P. Lanyon \textit{et al.}, Measurement-based quantum computation with trapped ions. \textit{Phys. Rev. Lett.} \textbf{111}, 210501 (2013).

\noindent [12] A. D. C\'{o}rcoles \textit{et al.}, Demonstration of a quantum error detection code using a square lattice of four superconducting qubits. \textit{Nat. Commun.} \textbf{6}, 6979 (2015).

\noindent [13] D. Rist\`{e} \textit{et al.}, Detecting bit-flip errors in a logical qubit using stabilizer measurements. \textit{Nat. Commun.} \textbf{6}, 6983 (2015).

\noindent [14] J. Kelly \textit{et al.}, State preservation by repetitive error detection in a superconducting quantum circuit. \textit{Nature} \textbf{519}, 66-69 (2015).

\noindent [15] L. Sun \textit{et al.}, Tracking photon jumps with repeated quantum non-demolition parity measurements. \textit{Nature} \textbf{511}, 444-448 (2014).

\noindent [16] G. D. Fuchs, V. V. Dobrovitski, D. M. Toyli, F. J. Heremans, D. D. Awschalom, Gigahertz dynamics of a strongly driven single quantum spin. \textit{Science} \textbf{326}, 1520-1522 (2009). 

\noindent [17] L. Robledo \textit{et al.}, High-fidelity projective read-out of a solid-state spin quantum register. \textit{Nature} \textbf{477}, 574-578 (2011).   

\noindent [18] N. Bar-Gill, L. M. Pham, A. Jarmola, D. Budker, R. Walsworth, Solid-state electronic spin coherence time approaching one second. \textit{Nat. Commun.} \textbf{4}, 1743 (2013).

\noindent [19] G.-Q. Liu, H. C. Po, J. Du, R.-B. Liu,  X.-Y. Pan, Noise-resilient quantum evolution steered by dynamical decoupling. \textit{Nat. Commun.} \textbf{4}, 2254 (2013).

\noindent [20] L. Jiang \textit{et al.}, Coherence of an optically illuminated single nuclear spin qubit.  \textit{Phys. Rev. Lett.} \textbf{100}, 073001 (2008).

\noindent [21] W. Pfaff \textit{et al.}, Demonstration of entanglement-by-measurement of solid-state qubits. \textit{Nature Phys.} \textbf{9}, 29-33 (2013).

\noindent [22] G. Wolfowicz \textit{et al.}, $^{29}$Si nuclear spins as a resource for donor spin qubits in silicon. arXiv:1505.02057 (2015).

\noindent [23] M. S. Blok \textit{et al.}, Manipulating a qubit through the backaction of sequential partial measurements and real-time feedback.  \textit{Nature Phys.} \textbf{10}, 189-193 (2014).

\noindent [24] J. T. Barreiro \textit{et al.}, An open-system quantum simulator with trapped ions. \textit{Nature} \textbf{470}, 486-491 (2011).

\noindent [25] D. Rist\`{e} \textit{et al.}, Deterministic entanglement of superconducting qubits by parity measurement and feedback. \textit{Nature} \textbf{502}, 350-354 (2013).

\noindent [26] P. London, \textit{et al.}, Detecting and polarizing nuclear spins with double resonance on a single electron spin. \textit{Phys. Rev. Lett.} \textbf{111}, 067601 (2013).

\noindent [27] M. D. Shulman \textit{et al.}, Suppressing qubit dephasing using real-time Hamiltonian estimation. \textit{Nat. Commun.} \textbf{5}, 5156 (2014).

\noindent [28] P. C. Maurer \textit{et al.}, Room-temperature quantum bit memory exceeding one second. \textit{Science} \textbf{336}, 1283-1286 (2012).

\noindent [29] G. Balasubramanian, \textit{et al.}, Ultralong spin coherence time in isotopically engineered diamond. \textit{Nature Mater.} \textbf{8}, 383 - 387 (2009).

\noindent [30] K. Saeedi \textit{et al.}, Room-temperature quantum bit storage exceeding 39 minutes using ionized donors in silicon-28. \textit{Science} \textbf{342}, 830-833 (2013).

\noindent [31] J. T. Muhonen \textit{et al.}, Storing quantum information for 30 seconds in a nanoelectronic device. \textit{Nature Nanotech.} \textbf{9}, 986–991 (2014).

\noindent [32] L. Li \textit{et al.}, Coherent spin control of a nanocavity-enhanced qubit in diamond. \textit{Nat. Commun.} \textbf{6}, 6173 (2015).

\noindent [33] F. Dolde \textit{et al.}, High-fidelity spin entanglement using optimal control. \textit{Nat. Commun.} \textbf{5}, 3371 (2014).

\noindent [34] W. Pfaff \textit{et al.}, Unconditional quantum teleportation between distant solid-state quantum bits. \textit{Science} \textbf{345}, 532-535 (2014).

\noindent [35] B. Hensen \textit{et al.}, Loophole-free Bell inequality violation using electron spins separated by 1.3 kilometres. \textit{Nature} \textbf{526}, 682-686 (2015).

\noindent [36] N. H. Nickerson, Y. Li, S. C. Benjamin, Topological quantum computing with a very noisy network and local error rates approaching one percent. \textit{Nat. Commun.} \textbf{4}, 1756 (2013).

\noindent [37] J. P. Hadden \textit{et al.}, Strongly enhanced photon collection from diamond defect centers under microfabricated integrated solid immersion lenses. \textit{Appl. Phys. Lett.} \textbf{97}, 241901 (2010). 

\noindent [38] T. K. Yeung \textit{et al.}, Anti-reflection coating for nitrogen-vacancy optical measurements in diamond. \textit{Appl. Phys. Lett.} \textbf{100}, 251111 (2012). 

\section{Acknowledgements}
We thank L. Dicarlo, L. M. K. Vandersypen, A. G. Fowler and V. V. Dobrovitski for discussions and comments. We acknowledge support from the Dutch Organization for Fundamental Research on Matter (FOM), the Netherlands Organization for Scientific Research (NWO), the Defense Advanced Research Projects Agency QuASAR program, the European Union S3NANO programs, and the European Research Council through a Starting Grant. THT is supported by an NWO VENI grant.

\newpage
\FloatBarrier

\begin{figure}[t]
\begin{center}
\includegraphics[scale=0.50]{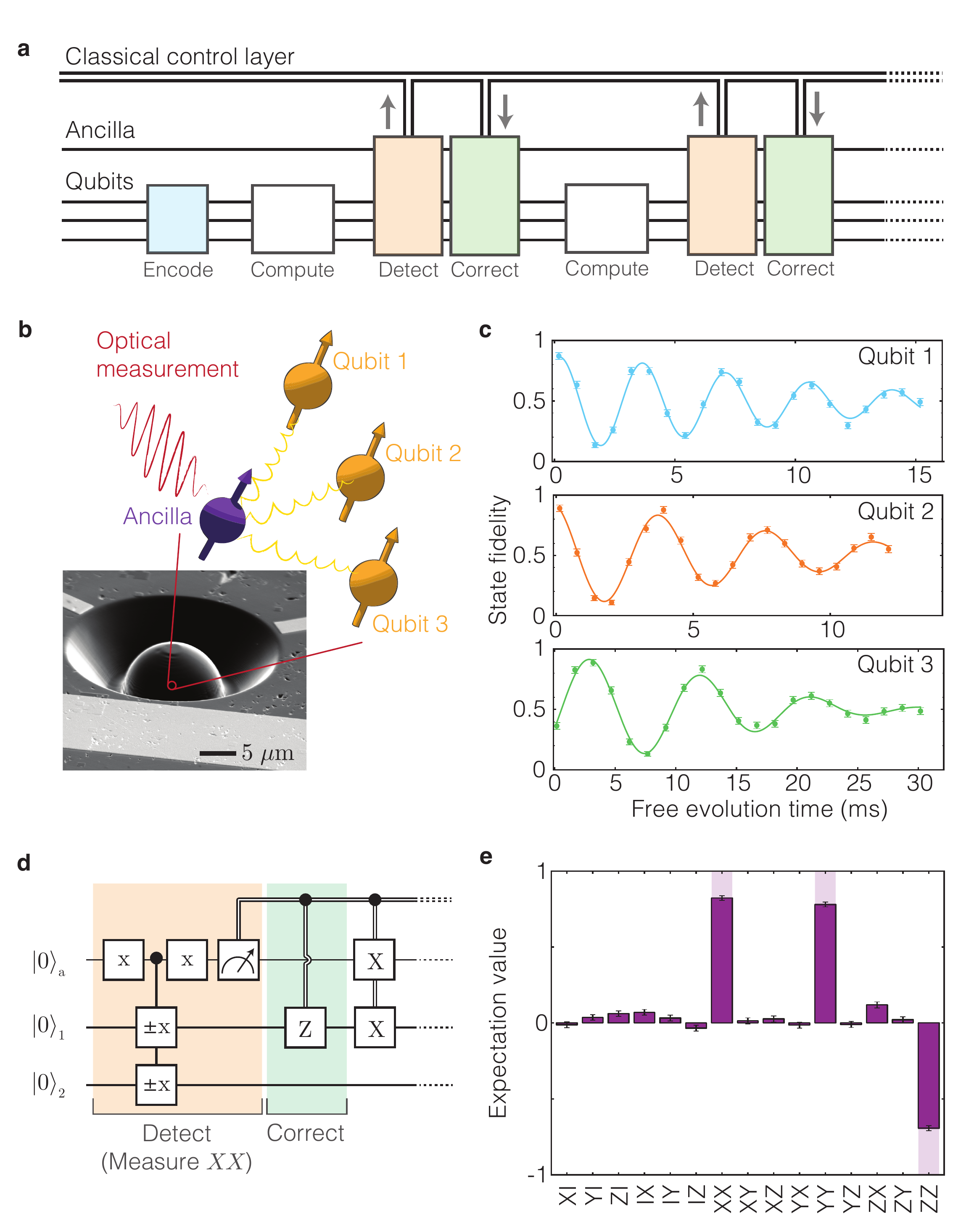}
\end{center}
\captionsetup{justification=raggedright}
\caption{\textbf{Quantum error correction and implementation of stabilizer measurements.} \textbf{a,} A quantum state is encoded in a logical qubit consisting of three physical qubits. Errors inevitably occur, for example during computations. An ancilla is used to repeatedly perform measurements that detect errors. Errors are corrected through classical logic and feedback, while the quantum state remains coherent and encoded. \textbf{b,} Device: CVD-grown single-crystal diamond with a solid-immersion lens [37] and on-chip lines for microwave control. Ancilla: the optically addressable electronic spin of a nitrogen vacancy (NV) centre. Qubits: three $^{13}$C nuclear spins that are controlled and measured through the hyperfine coupling to the ancilla (Methods). \textbf{c,} Free induction decay (Ramsey) experiments. Gaussian fits yield dephasing times $T_2^* =$ = 12.0(9), 9.1(6) and 18.2(9) ms for qubits 1, 2 and 3, respectively. \textbf{d,} Deterministic entanglement of two qubits by $XX$ stabilizer measurement and feedback. The $\pm x$ gates are $\pi/2$ rotations around $x$ with the sign controlled by the ancilla state. The final  operations reset the ancilla and account for an additional  flip for the +1 outcome (Methods). \textbf{e,} State tomography of the generated entangled state for qubits 2 and 3. The fidelity with the ideal state is $F=0.824(7)$ (see Supplementary Fig. 6 for other qubit combinations and post-selected results). } \label{Figure1}
\vspace*{-0.2cm}
\end{figure}

\begin{figure}[t]
\begin{center}
\includegraphics[scale=0.50]{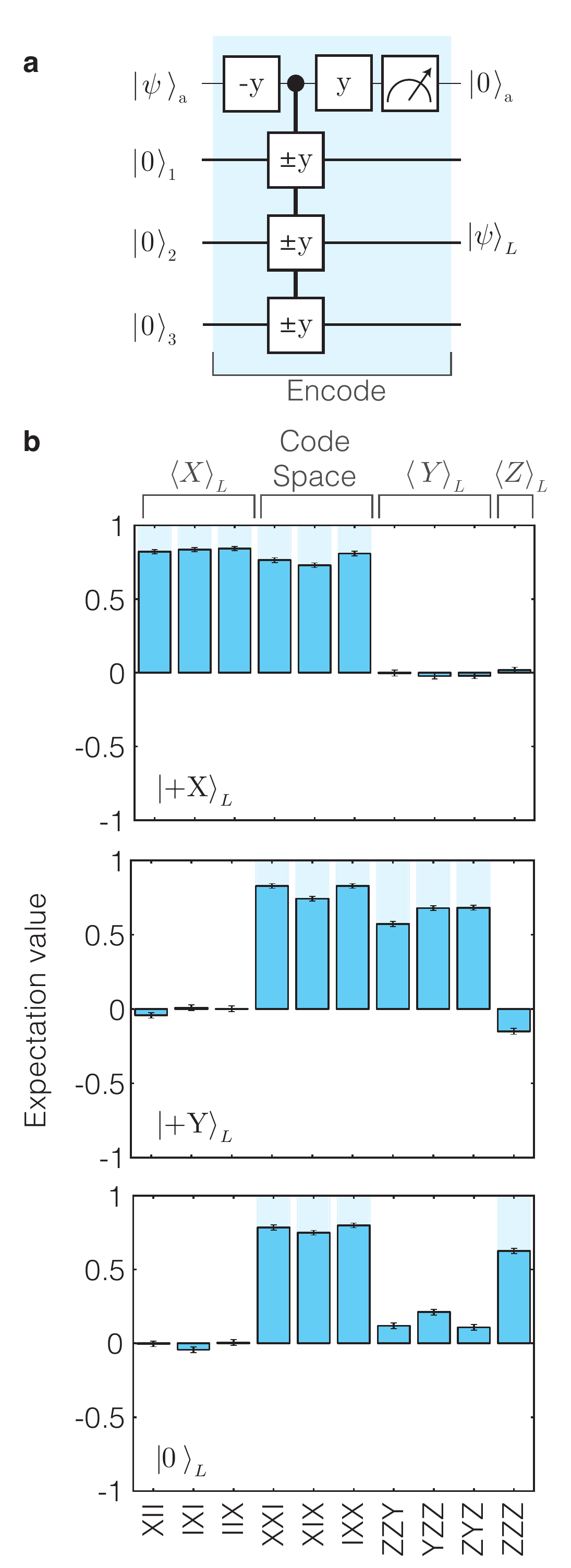}
\end{center}
\captionsetup{justification=raggedright}
\caption{\textbf{Encoding of the logical qubit.} \textbf{a,} 
Encoding an arbitrary quantum state $\ket{\psi}=\alpha\ket{0}+\beta\ket{1}$ prepared on the ancilla into $\ket{\psi}_L=\alpha\ket{0}_L+\beta\ket{1}_L$. Successful encoding is heralded by outcome $\ket{0}_a$. \textbf{b,}  Characterization of the logical states 
$\ket{+X}_L$, $\ket{+Y}_L$ and $\ket{0}_L$. Only the logical qubit operators and stabilizers are shown (see Supplementary Fig. 7 for complete tomography of all 6 logical basis states). The fidelities with the ideal three-qubit states are $F=0.810(5)$, $0.759(5)$ and $0.739(5)$, respectively, 
demonstrating three-qubit entanglement [10]. The logical state fidelities are $F_{+X}=(1+\ev{X_L})/2=0.916(6)$, $F_{+Y}=(1+\ev{Y_L})/2=0.822(7)$ and $F_0=(1+\ev{Z_L})/2=0.813(9)$.  Ideally, all the encoded states are +1 eigenstates of the stabilizers $X_1X_2I_3$ and $I_1X_2X_3$. The fidelity to this code space, $F_S=(1+\ev{X_1X_2I_3}+\ev{I_1X_2X_3}+\ev{X_1I_2X_3})/4$, is $0.839(3)$ averaged over all states and gives the probability that the starting state is free of detectable errors. 
} \label{Figure2}
\vspace*{-0.2cm}
\end{figure}

\begin{figure}[t]
\begin{center}
\includegraphics[scale=0.50]{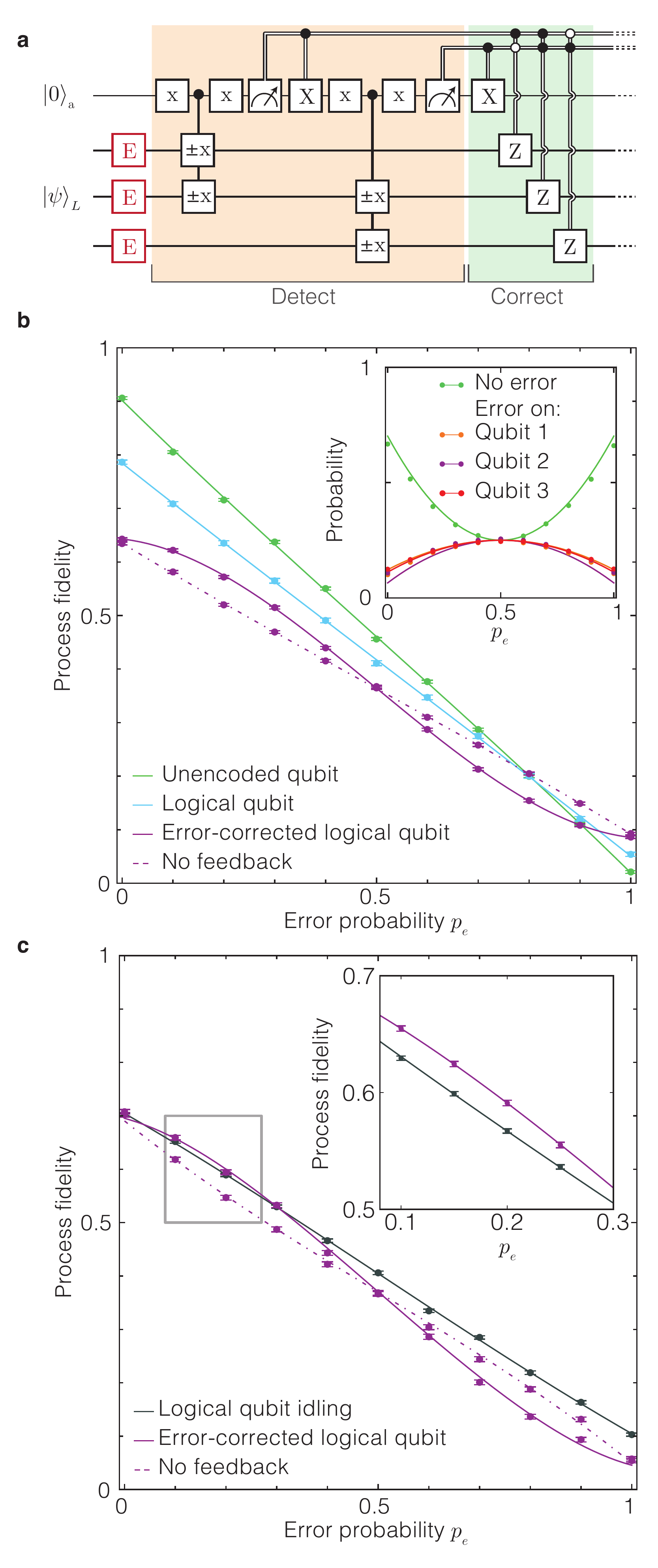}
\end{center}
\captionsetup{justification=raggedright}
\caption{\textbf{Active quantum error correction by stabilizer measurements.} \textbf{a,} 
All qubits are simultaneously subjected to uncorrelated phase errors $E$ with probability $p_e$. Errors are detected by measuring $X_1X_2I_3$ and $I_1X_2X_3$ and subsequently corrected by $Z$ operations through feedback. Finally we measure the process fidelity with the identity. 
\textbf{b,} Process fidelities for: an unencoded qubit, the logical qubit without stabilizer measurements, the error-corrected logical qubit, and the logical qubit without feedback (i.e. errors are detected but not corrected). We average over the logical qubit permutations, e.g. $X_L = X_1I_2I_3$, 
$I_1X_2I_3$ and $I_1I_2X_3$, and the four ways to assign the ancilla states to the error syndromes (see Supplementary Fig. 8 for individual curves). Inset: probabilities for the error syndromes with theoretically predicted curves based on the state tomography in Fig. 2b (Supplementary Note 2).  
\textbf{c,} Comparison between the error-corrected logical qubit and the logical qubit with the stabilizer measurements replaced by an equivalent idle time (2.99 ms). Compared to b, the effective readout fidelity is optimized by associating syndrome +1,+1 (no error) to obtaining $\ket{1}_a$ for both stabilizer measurements. Curves in \textbf{b} and \textbf{c} are fits described in the Methods.}
\label{Figure3}
\vspace*{-0.2cm}
\end{figure}

\begin{figure}[t]
\begin{center}
\includegraphics[scale=0.50]{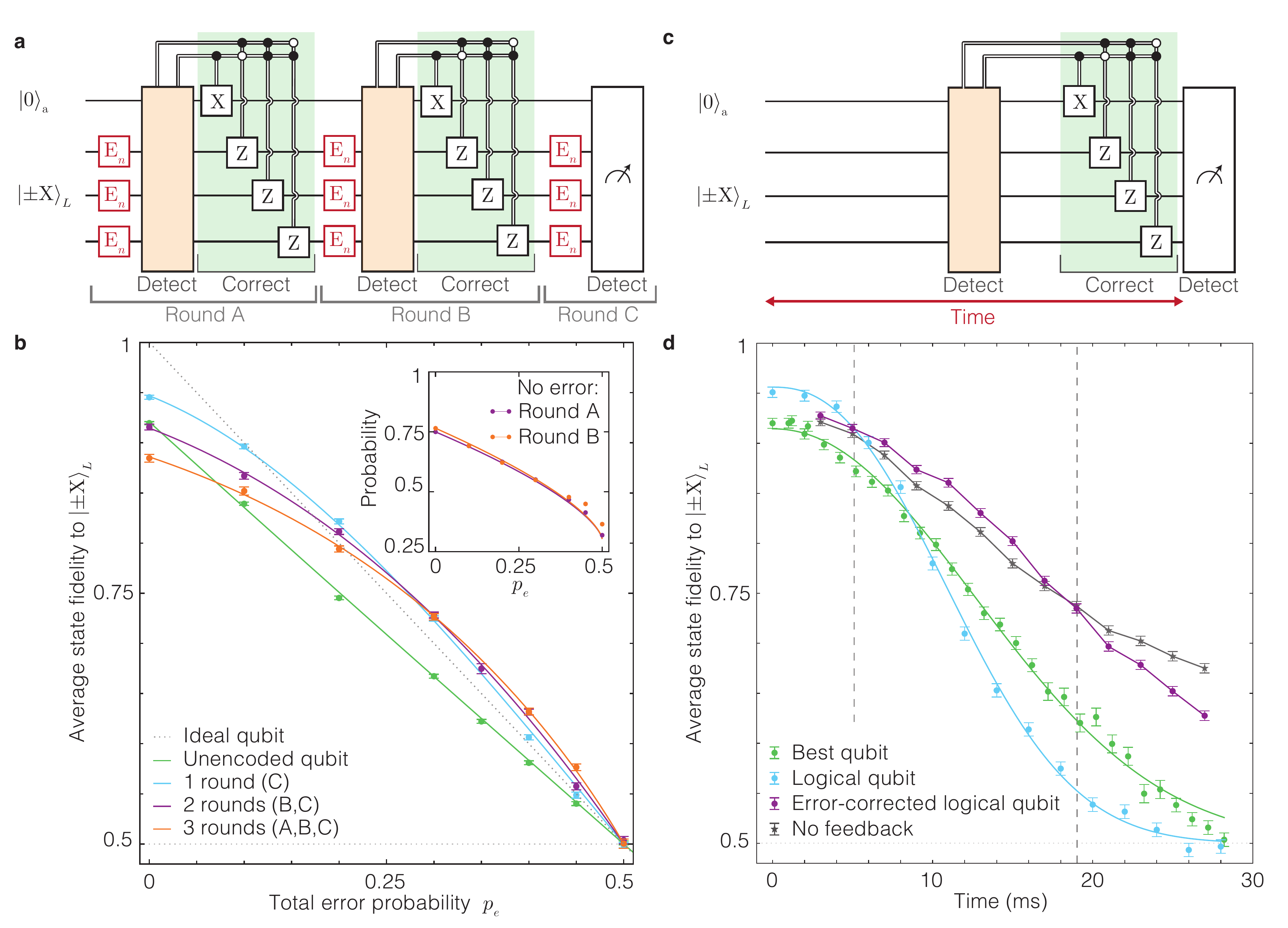}
\end{center}
\captionsetup{justification=raggedright}
\caption{\textbf{ Extending coherence by active quantum error correction. a,} 
Three rounds of error correction on a logical qubit. The first two rounds of quantum error correction use stabilizer measurements and feedback. The final round is implemented by majority voting. \textbf{b,} Average logical state fidelity for $\ket{+X}_L$ and $\ket{-X}_L$ as a function of total error probability $p_e$ for $n$ = 1, 2 and 3 rounds of error correction compared to an unencoded qubit. The errors per round $E_n$ occur with probability $p_n$. Inset: probabilities that no error is detected ($n=3$). The similarity of the results for rounds A and B confirms that errors are corrected in between rounds. \textbf{c,} Correcting natural dephasing. The storage time is defined from the end of the encoding until the start of the final measurements. \textbf{d,} Dephasing of the logical qubit: without stabilizer measurements, with quantum error correction and without feedback, compared to the best unencoded qubit. The dashed lines indicate the times between which the actively error-corrected logical qubit gives the highest fidelity. The data without feedback (detecting errors without correcting) isolates the suppression of coherently evolving errors by projecting them. For long times, applying error correction lowers the fidelity because the stabilizer measurements extract no useful information about errors, but nevertheless preferentially suppress evolutions that result in phase errors at the end of the sequence (See Supplementary Fig. 10 for a detailed analysis). See Supplementary Fig. 9 for error syndrome probabilities. Curves in \textbf{b} and \textbf{d} are fits described in the Methods and Supplementary Notes 1\&2.} \label{Figure4}
\vspace*{-0.2cm}
\end{figure}

\FloatBarrier
\newpage
\section{Supplementary Information}

\beginsupplement

\subsection{Supplementary Note 1: Theoretical analysis: state and process fidelities}

For ideal error correction the process fidelity to the identity as a function of error probability $p_e$ for a single round of quantum error correction (QEC) is
\be
F_{QEC}(p_e)=O+A(1-3p_e^2+2p_e^3).
\label{Eq:FQEC}
\ee
The offset $O$ and amplitude $A$ account for the finite experimental state fidelities. Note that the value at $p_e = 0.5$, $F_{QEC} (p_e=0.5)= O+A/2$, is determined by the fidelity of the logical states $\ket{0}_L$ and $\ket{1}_L$, which are insensitive to phase errors.
Without error correction a linear function 
\be
F_{linear} = O+A(1-p_e)
\label{Eq:FN}
\ee
is expected. The experimental data can be fitted to a weighted sum of the two functions by:
\be
F_P = wF_{QEC}+(1-w)F_{linear}.
\label{Eq:QEC}
\ee
The shape of the curve is set by $w$, which gives the relative weights of the equations for ideal error correction and for no error correction.

\subsubsection{Assignment of ancilla states to the error syndrome: effective measurement fidelity $F_{\text{M}}$}

In our experiment, the ancilla readout fidelity is asymmetric: $\ket{1}_a$ has a higher readout fidelity ($F_1=0.988(2)$) than $\ket{0}_a$ ($F_0=0.890(4)$). The effective measurement fidelity for error correction $F_{\text{M}}$ is therefore determined by the probabilities to obtain $\ket{0}_a$ or $\ket{1}_a$, which depend on the assignment of the ancilla states ($\ket{0}_a$ or $\ket{1}_a$) to each stabilizer measurement outcome (+1 or -1) and the probabilities for different errors to occur. There are four different ways to assign the ancilla states to the error syndromes: the +1,+1 outcome (no error) can be set to result in $\{\ket{0}_a,\ket{0}_a\}$, $\{\ket{0}_a,\ket{1}_a\}$, $\{\ket{1}_a,\ket{0}_a\}$ or $\{\ket{1}_a,\ket{1}_a\}$. The probability to obtain outcome +1,+1 (no error) ideally is $1-3p_e+3p_e^2$, while 
the probability to detect an error on a given qubit is $p_e-p_e^2$. 
With these probabilities, we obtain the effective QEC measurement fidelity as function of error probability:
\be
F_{\text{M}} = F^{\text{(0)}}(1-3p_e+3p_e^3)+(F^{\text{(1)}}+F^{\text{(2)}}+F^{\text{(3)}})(p_e-p_e^2)=F^{\text{(0)}}+(F^{\text{(1)}}+F^{\text{(2)}}+F^{\text{(3)}}-3F^{\text{(0)}})(p_e-p_e^2),
\label{Eq:ED}
\ee
with $F^{(0)}$, $F^{(1)}$, $F^{(2)}$ and $F^{(3)}$, the readout fidelities for 0 errors, an error on qubit 1, an error on qubit 2 and an error on qubit 3, respectively. For example, for assignment $\{\ket{1}_a,\ket{1}_a\}$ to stabilizer outcomes +1,+1 (no error), these readout fidelities are
\begin{align*}
F^{\text{(0)}} &= F_1^2,\\
F^{\text{(1)}} &=F^{\text{(3)}} = F_1F_0,\\
F^{\text{(2)}}& = F_0^2,
\end{align*}
In a similar way, the fidelities for the other three assignments can be calculated.
\\ \\
Finally, if we assume that an erroneous ancilla readout decoheres the logical state, the dependence of the effective readout fidelity on $p_e$ can be taken into account by setting:
\be
A = A' F_{\text{M}}
\label{Eq:p_dependence_of_A}
\ee
in Eqs.~\ref{Eq:FQEC}$\&$\ref{Eq:FN} for the process fidelity, with $A'$ a constant.

\subsubsection{Fitting of Figs. 3b, 3c and \ref{Fig:Syn}}

In Fig. 3b, the ancilla readout is symmetrized by averaging over all four assignments, so that $F_{\text{M}}$ equals the average readout fidelity 0.939(2) and is independent of $p_e$. We can therefore simply fit the data in Fig. 3b to Eq. \ref{Eq:QEC}, with $A$ constant. We find $w = 0.81(3)$, corresponding to an average probability to successfully correct single-qubit errors of $\ev{P_n} = \frac{1}{3}(w+2)=0.94(1)$ [1]. We obtain $A=0.557(2)$ and $O=0.086(1)$. For the unencoded qubit, the encoded qubit without stabilizer measurements, and the encoded qubit without feedback, we find a linear function and $w \approx 0$ ($\ev{P_n} \approx 2/3$) as expected without error correction (exact values: $w = -0.06(3), -0.03(3)$ and $-0.07(3)$,  $A=0.882(4), 0.734(3)$ and $0.543(2)$, and $O=0.019(3),0.051(3)$ and $0.092(1)$, respectively for the three cases). In Fig.~\ref{Fig:Syn} the separate process fidelities for the different assignments are shown. Switching between assignments is done by adding or omitting a $\pi$-pulse before the ancilla readout. 
\\ \\
In Fig. 3c, we assign the ancilla state $\ket{1}_a$ to the +1 outcome for all stabilizer measurements. This assignment is optimal because it associates the best readout fidelity with the most likely outcome: +1,+1 (no error, inset in Fig.~3b). We fit the data in Fig. 3c and Fig.~\ref{Fig:Syn} to Eq.~\ref{Eq:QEC}, with $A$ now error-dependent according to Eq.~\ref{Eq:p_dependence_of_A} and obtain $w=0.8(1)$, corresponding to $\ev{P_n} = 0.93(3)$ ($A'=0.666(8)$ and $O=0.038(6)$). The values for $w$ and $\ev{P_n}$ are in good agreement with the result of Fig. 3b, indicating that the treatment in Eqs.~\ref{Eq:ED}\&\ref{Eq:p_dependence_of_A} is accurate.


\subsubsection{Multiple rounds of error correction (incoherent errors), Fig. 4b}
For multiple rounds of QEC with incoherent errors and with the total error with probability $p_e$ equally distributed over $n$ rounds, the error-probability per round is
$p_{n}=\frac{1}{2}(1-\sqrt[n]{1-2p_e})$, for $p_e<0.5$.
Ideally, the (average) state fidelity is then described by:
\be
F = \frac{1}{2}\lbrack1+(1-6p_n^2+4p_n^3)^n\rbrack.
\ee
As before we fit the data to a weighted sum of the equations for ideal error correction and for a linear error-dependence (no error correction). We use the optimal ancilla state assignment ($F_{\text{M}}(p_e)$ from Eq.~\ref{Eq:ED}).
For two rounds of error correction we obtain
\be
F_{2} =\frac{1}{2}w\lbrack1+A'F_{\text{M}}(1-6p_{2}^2+4p_{2}^3)^2\rbrack+\frac{1}{2}(1-w)\lbrack1+A'F_{\text{M}}(1-2p_e)\rbrack,
\ee
giving $w = 0.66(4)$ and $A' = 0.850(9)$.
For three rounds it becomes
\be
F_{3} =\frac{1}{2}w\lbrack1+A'F^2_{\text{M}}(1-6p_{3}^2+4p_{3}^3)^3\rbrack+\frac{1}{2}(1-w)\lbrack1+A'F^2_{\text{M}}(1-2p_e)\rbrack,
\ee
giving $w = 0.71(2)$ and $A' = 0.810(5)$. Importantly, the data for multiple rounds cannot be accurately described by the expected shape for a single round of error correction (Eq. \ref{Eq:QEC}).

\subsubsection{Naturally occurring decoherence (coherent errors), Fig. 4d}
The experiments for the best unencoded qubit and logical qubit (majority vote only) are fitted to a general exponentially decaying function:
\be
F = \frac{1}{2}(1+Ae^{-(t/T)^n}).
\label{Eq:decay}
\ee
Here, we obtain for the best qubit: $T=17.3(2)$ ms and $n=2.09(7)$, while for the encoded qubit with majority voting we obtain: $T=13.7(1)$ ms and $n=2.37(8)$. 
\\ \\
To understand the interplay of quantum error correction and the projection of errors in the experiments with stabilizer measurements at half the free evolution time in Fig.~4d we turn to numerical Monte Carlo simulations, see Fig.~\ref{Fig10} for details and results.

\subsection{Supplementary Note 2: Theoretical analysis: error probabilities}
The probability to detect no error ($P^{(0)}$) is the sum of the probability to have no error (no qubits flipped) or three errors (all qubits flipped) and is described by:
\be
P^{(0)}=(1-p_{\text{tot}}^{\text{(1)}})(1-p_{\text{tot}}^{\text{(2)}})(1-p_{\text{tot}}^{\text{(3)}})+p_{\text{tot}}^{\text{(1)}}p_{\text{tot}}^{\text{(2)}}p_{\text{tot}}^{\text{(3)}},
\label{Eq:no}
\ee
where $p_{\text{tot}}^{(i)}$ is the error probability for qubit $i$.
The probability to detect an error on one of the three qubits ($P^{(i)}$) is the probability to have an error on qubit $i$, or an error on both of the other qubits, which for example for qubit 1 is described by:
\be
P^{(1)} =p_{\text{tot}}^{\text{(1)}}(1-p_{\text{tot}}^{\text{(2)}})(1-p_{\text{tot}}^{\text{(3)}})+(1-p_{\text{tot}}^{\text{(1)}})p_{\text{tot}}^{\text{(2)}}p_{\text{tot}}^{\text{(3)}}.
\label{Eq:E}
\ee
With finite input error probability $p_{\text{in}}^{(i)}$ for qubit $i$ (errors already present in the initially prepared state), the total error as function of the applied error probability $p_e$, becomes:
\be
p_{\text{tot}}^{(i)}=p_{\text{in}}^{(i)}+p_e-2p_{\text{in}}^{(i)}p_e
\label{Eq:ptot}
\ee
\\
Finally we can take imperfect ancilla readout into account and obtain the probability to detect one of the error outcomes $P_{\text{D}}^{(i)}$ ($i$=0 for no detected error) as function of the applied error $p_e$:
\begin{align}
P_{\text{D}}^{(0)}&=  P^{(0)}F^2+ (P^{(1)}+P^{(3)})F(1-F)+P^{(2)}(1-F)^2\label{Eq:PD0} \\
P_{\text{D}}^{(1)} &= P^{(1)}F^2+ (P^{(0)}+P^{(2)})F(1-F)+P^{(3)}(1-F)^2\\
P_{\text{D}}^{(2)} &= P^{(2)}F^2+ (P^{(1)}+P^{(3)})F(1-F)+P^{(0)}(1-F)^2\\
P_{\text{D}}^{(3)} &= P^{(3)}F^2+ (P^{(0)}+P^{(2)})F(1-F)+P^{(1)}(1-F)^2\label{Eq:PD3}
\end{align}
\\ \\
The $XX$ stabilizers in the encoded state tomography (Fig.~2) detect errors present in the encoded state, we obtain:
\begin{align*}
P^{(0)} =&\ev{\frac{1}{4}(1+X_1,X_2,I_3)(1+X_1,I_2,X_3)}=\frac{1}{4}(1+\ev{X_1,X_2,I_3}+\ev{I_1,X_2,X_3}+\ev{X_1,I_2,X_3}) = 0.785(2)\\
P^{(1)}=&\ev{\frac{1}{4}(1-X_1,X_2,I_3)(1-X_1,I_2,X_3)}=\frac{1}{4}(1-\ev{X_1,X_2,I_3}+\ev{I_1,X_2,X_3}-\ev{X_1,I_2,X_3}) = 0.060(2)\\ 
P^{(2)}=&\ev{\frac{1}{4}(1-X_1,X_2,I_3)(1+X_1,I_2,X_3)}=\frac{1}{4}(1-\ev{X_1,X_2,I_3}-\ev{I_1,X_2,X_3}+\ev{X_1,I_2,X_3}) = 0.083(2)\\ 
P^{(3)} =&\ev{\frac{1}{4}(1+X_1,X_2,I_3)(1-X_1,I_2,X_3)}=\frac{1}{4}(1+\ev{X_1,X_2,I_3}-\ev{I_1,X_2,X_3}-\ev{X_1,I_2,X_3}) = 0.071(2) 
\end{align*}
which are uncorrected for qubit readout.
These results can be translated to the input errors, as these outcomes refer to Eqs.~\ref{Eq:no}\&\ref{Eq:E} with no additional applied error $p_e$, giving $p_{\text{in}}^{(1)} = 0.064(2)$, $p_{\text{in}}^{(2)} = 0.091(2)$, $p_{\text{in}}^{(3)} = 0.077(2)$.
\\ \\
Using these values we estimate the expected total error detection probabilities $P_{\text{D}}^{(0)}$,  $P_{\text{D}}^{(1)}$,  $P_{\text{D}}^{(2)}$ and  $P_{\text{D}}^{(3)}$ as function of applied error probability $p_e$ according to Eqs.~\ref{Eq:no}-\ref{Eq:PD3}.
The expected error-dependent QEC measurement outcomes are shown by the solid lines in the inset of Fig.~3b.

\subsubsection{Error syndrome assignment}
For the different error assignments, the asymmetry in the ancilla readout complicates the error detection curves: the QEC measurement fidelity is dependent on the error probability.
If, for instance, both stabilizer measurements giving +1 are assigned to $\{\ket{1}_a,\ket{1}_a\}$, Eqs.~\ref{Eq:PD0}-\ref{Eq:PD3} become:
\begin{align}
P_{11}^{(0)} &= P^{(0)}F_{\text{1}}^2+ (P^{(1)}+P^{(3)})F_{\text{1}}(1-F_{\text{0}})+P^{(2)}(1-F_{\text0})^2 \label{Eq:no_11} \\
P^{(1)}_{11} &= P^{(1)}F_{\text{1}}F_{\text{0}}+P^{(0)}F_{\text{1}}(1-F_{\text{1}})+P^{(2)}F_{\text{0}}(1-F_{\text{0}})+P^{(3)}(1-F_{\text{1}})(1-F_{\text{0}})\label{Eq:1_11} \\
P^{(2)}_{11} &= P^{(2)}F_{\text{0}}^2+ (P^{(1)}+P^{(3)})F_{\text{0}}(1-F_{\text{1}})+P^{(0)}(1-F_{\text1})^2\label{Eq:2_11}  \\
P^{(3)}_{11} &= P^{(3)}F_{\text{1}}F_{\text{0}}+P^{(0)}F_{\text{1}}(1-F_{\text{1}})+P^{(2)}F_{\text{0}}(1-F_{\text{0}})+P^{(2)}(1-F_{\text{1}})(1-F_{\text{0}}\label{Eq:3_11} )
\end{align}
All error detection curves for the four error assignments using similar equations are plotted in Fig.~\ref{Fig:Syn}.
\subsubsection{Multiple rounds of error correction, Fig. 4b}
For multiple rounds we now calculate the average input error $p_{\text{in}}^{\text{(avg)}}$ from the detection probability for no additional applied error ($p_e = 0 $).
We simplify Eq.~\ref{Eq:no} to
\be
P^{(0)} = 1-3p_{\text{tot}}^{(\text{avg})}+3(p_{\text{tot}}^{(\text{avg})})^2
\ee
and use Eq.~\ref{Eq:no_11} to obtain the following average input error for  round 1: $p_{\text{in}}^{\text{(avg)}} = 0.092(1)$ and for round 2: $p_{\text{in}}^{\text{(avg)}}= 0.086(1)$.
The resulting curves according to Eq.~\ref{Eq:no_11} are shown in the inset of Fig. 4b.

\subsection{Supplementary Note 3: Qubit readout calibration}
To obtain best estimates for the actual states, the results are corrected for the fidelity of the gates used in the final readout (tomography). We distinguish between reading out single- two- and three-qubit expectation values. 
\\ \\
For a single qubit $i$ that is initialized and readout immediately, the measured expectation value $\ev{Z_i}$ is set by the initialization fidelity of the nitrogen spin ($F_N = 0.94(3)$) and by factors due to the initialization ($C_{\text{init},Qi}$) and readout ($C_{Q_i}$) of the qubit. Because the initialization and readout consist of the same set of gates, we assume that $C_{\text{init},Qi} = C_{Q_i}$ for this experiment and obtain: 
\begin{equation}
\ev{Z_i}=F_NC^2_{Q_i}, 
\label{Eq:1Qev}
\end{equation}
from which a readout correction factor $1/C_{Q_i}$ can be determined.
\\ \\
To calibrate the multi-qubit readouts we initialize the three qubits in separable states. For example, for state $\ket{000}$, the measured three-qubit expectation value $\ev{Z_1Z_2Z_3}$ is set by the nitrogen initialization $F_N$, by factors $C_{\text{init},Qi}$ due to the individual initialization fidelities of the three-qubits and by a factor $C_{Q_1,Q_2,Q_3}$ due to the three-qubit readout:
\be
\ev{Z_1Z_2Z_3}=F_NC_{\text{init},Q1}C_{\text{init},Q2}C_{\text{init},Q3}C_{Q_1,Q_2,Q_3} \to C_{Q_1,Q_2,Q_3} = \frac{\ev{Z_1Z_2Z_3}}{F_NC_{\text{init},Q1}C_{\text{init},Q2}C_{\text{init},Q3}}.
\ee
This equation assumes that the initialization errors, other than those due to the nitrogen initialization, are uncorrelated. The initialization fidelities are obtained using the single-qubit expectation values and  single qubit $C_{Q_i}$ for the corresponding qubit, i.e. for qubit 1:
\be
\ev{Z_1I_2I_3}=F_NC_{\text{init},Q1}C_{Q_1} \to C_{\text{init},Q1} = \frac{\ev{Z_1I_2I_3}}{F_N C_{Q_1}}
\ee
with $C_{Q_1}$ from Eq.~\ref{Eq:1Qev}. In a similar way, the two-qubit readout is calibrated using two-qubit expectation values of two- and three-qubit states.
We obtain the following values:
\begin{align*}
C_{Q_1}&=0.95(1) &C_{Q_1Q_2}&= 0.94(2)&C_{Q_1Q_2Q_3}&= 0.92(5)\\
C_{Q_2}&=0.94(1) &C_{Q_1Q_3}&= 0.88(4)\\
C_{Q_3}&=0.95(1) &C_{Q_2Q_3}&= 0.90(2)
\end{align*}
which are used to calibrate the final readouts for tomography. Note that the uncertainty in the readout calibration potentially creates a small systematic error (a rescaling of all y-axes). For this reason we also provide all raw (uncalibrated) data for the error correction  in Fig.~\ref{Fig:fig3_uncorrected}.   

\newpage
\section{Supplementary Tables}
\renewcommand{\tablename}{Supplementary Table}

\begin{table}[h!]
  \centering
   \begin{tabular}{c|c|c|c}
        				& Qubit 1						& 	          Qubit 2 	 		     &	Qubit 3 	\\[1mm]
      \hline
      $A_\parallel$ (kHz) 	& $\ \  $ 2$\pi\cdot$20.6 $\ \ $ 	& $\ \ $ 2$\pi\cdot$-36.4$\ \ $&$\ \ $  2$\pi\cdot$24.4 \\
      $A_\perp$ (kHz)     	& $\ \  $ 2$\pi\cdot$43   $\ \ $ 	& $\ \ $ 2$\pi\cdot$25    $\ \ $&$\ \ $  2$\pi\cdot$26 \\
      $\omega_0$ (kHz)   & $\ \  $ 2$\pi\cdot$431.874(3)           $\ \ $ 	& $\ \ $ 2$\pi\cdot$431.994(3)              $\ \ $&$\ \ $  2$\pi\cdot$431.934(3) \\
      $\omega_1$ (kHz)   & $\ \  $ 2$\pi\cdot$413.430(3)           $\ \ $ 	& $\ \ $ 2$\pi\cdot$469.025(3)             $\ \ $&$\ \ $  2$\pi\cdot$408.303(3) \\
      $\tau$ ($\mu$s) 	& $\ \  $ 13.616                   $\ \ $ 	& $\ \ $ 4.996           	    $\ \ $&$\ \ $  11.312 \\
      N  				& $\ \  $ 32                   	     $\ \ $ 	& $\ \ $ 34      	    		    $\ \ $&$\ \ $ 48 \\
      gate time ($\mu$s) 	& $\ \  $ 980                  	     $\ \ $ 	& $\ \ $ 400      	           $\ \ $&$\ \ $ 1086 \\
      $T_2^*$, $m_s=\ \ 0$ (ms) & $\ \  $ 12.0(9)        $\ \ $ 	& $\ \ $ 9.1(6)      	    $\ \ $&$\ \ $ 18.2(9) \\
      $T_2^*$, $m_s=-1$ (ms) & $\ \  $ 12.8(6)           $\ \ $ 	& $\ \ $ 9.8(4)      	           $\ \ $&$\ \ $ 21(1) \\
      $T_1$, $m_s=\ \ 0$ (ms) & $\ \  $ 110(10)       $\ \ $ 	& $\ \ $ 100(10)     	    $\ \ $&$\ \ $ 330(30) \\
   \end{tabular}
\captionsetup{justification=raggedright}
  \caption{\textbf{Qubit and gate parameters.}  $A_\parallel$ and $A_\perp$ are the estimated hyperfine interaction components parallel and perpendicular to the applied magnetic field. $\omega_{0}$ and $\omega_{1}$ are the nuclear precession frequencies for $m_s=0$ ($\ket{0}_a$) and $m_s=-1$ ($\ket{1}_a$). $\tau$ is half the inter pulse delay, $N$ the number of pulses and $gate\ time$ the total duration for the conditional $\pm x$-gates. These values vary slightly over the experiment as they are calibrated every $\sim36$ hours. $T_2^*$ is the (natural) dephasing time and $T_1$ the longitudinal relaxation time.
}\label{Tab:Qubits}
\end{table}

\begin{table}[h!]
  \centering
   \begin{tabular}{l|c|c}
        				& 						          Fig. 3b (logical qubit with QEC)    		     &	 Fig. 4b (three rounds)	\\[1mm]
      \hline
       Two-qubit gates 	& 19 & 20\\
       Ancilla refocussing pulses & 698 & 808 \\
       Ancilla read-out and reset & 7 & 9\\
   \end{tabular}
\captionsetup{justification=raggedright}
  \caption{\textbf{Experimental complexity.} Number of operations in the sequence starting from the initialization of the qubits. All qubit ($^{13}$C) gates are composed of ancilla (NV electron spin) refocussing pulses and the ancilla is read-out and reset multiple times. We give values for two examples: a single round of QEC with measurement of $\ev{Z_1Z_2Z_3}$ (Fig. 3b) and three rounds of QEC with measurement of $\ev{X_1X_2X_3}$ (Fig. 4b).
}\label{Tab:Details}
\end{table}

\newpage
\section{Supplementary Figures}

\renewcommand{\figurename}{Supplementary Figure}
\begin{figure}[h!]
\begin{center}
\includegraphics[scale=0.25]{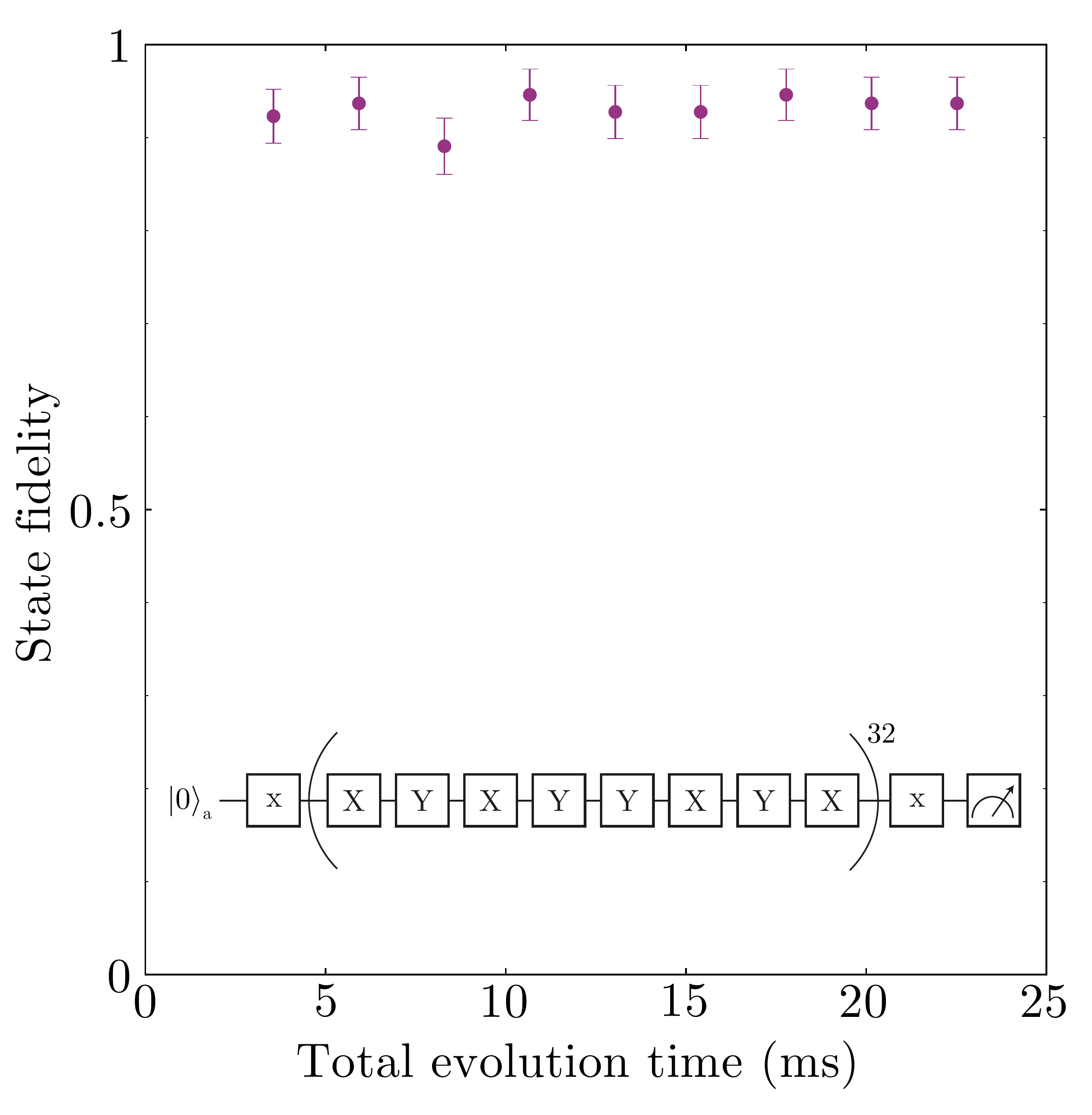}
\end{center}
\captionsetup{justification=raggedright}
\caption{\textbf{Coherence of the NV electron spin by dynamical decoupling with 256 pulses.} 
The spin is decoupled from the nuclear spin bath by applying a sequence of 256 $\pi$-pulses with alternating phases. The time between the pulses is chosen to be a multiple of the Larmor period of the $^{13}$C spins. This result shows no significant decay on the relevant timescale of our experiments.}
\label{Fig:DD} \vspace*{-0.2cm}
\end{figure} 

\begin{figure}[h!]
\begin{center}
\includegraphics[scale=0.55]{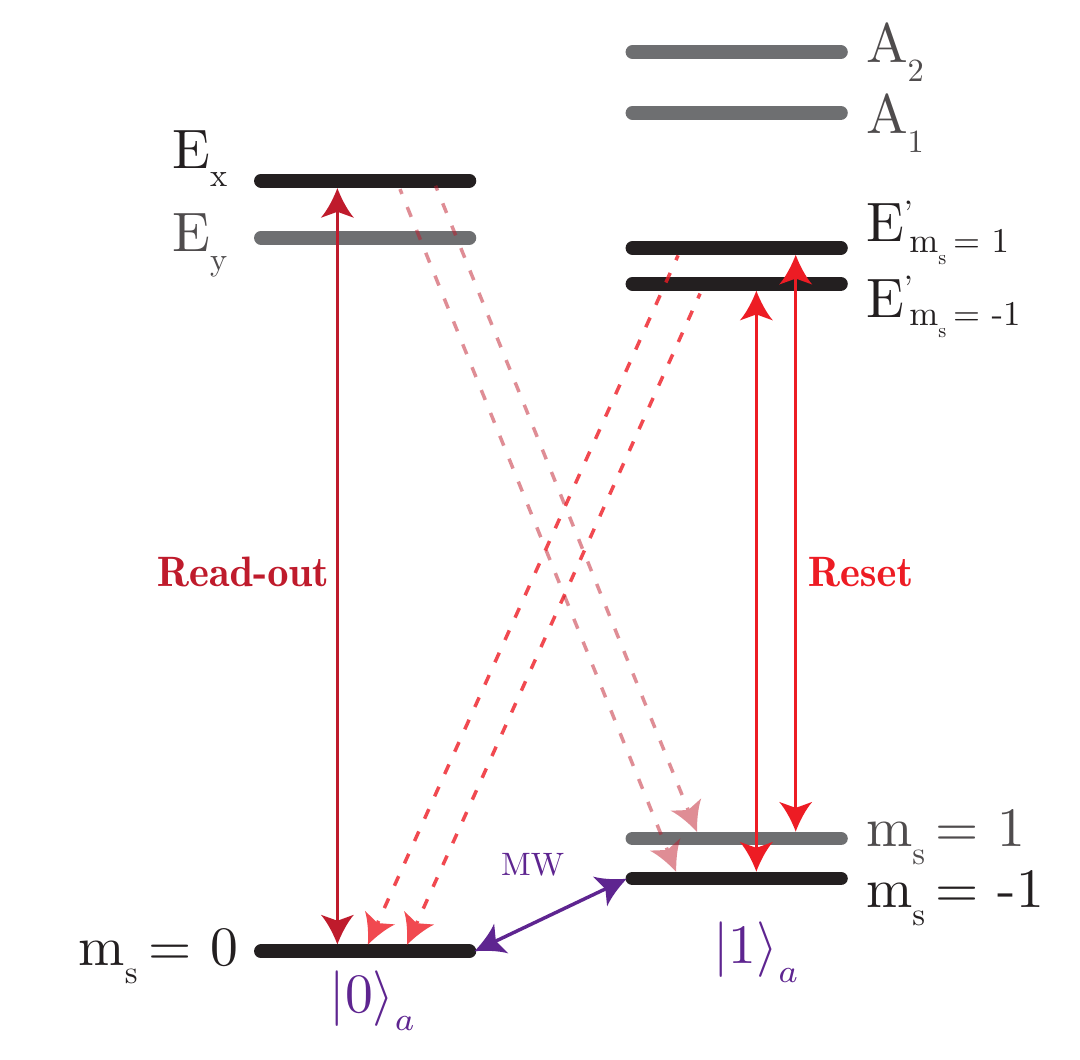}
\end{center}
\captionsetup{justification=raggedright}
\caption{\textbf{Optical readout and initialization of the ancilla NV electron spin.} The electron spin is initialized and read out by spin-selective resonant excitation [2]. To initialize or reset the electron spin state we apply a laser pulse that excites only the $m_s=\pm1 \leftrightarrow E'_{m_s=\pm1}$ transitions (Reset). Due to spin mixing in the excited state this prepares the electron spin in the $m_s=0$ state (fidelity $> 0.98$) [2].
To measure the spin state we apply a laser pulse resonant with the $m_s=0 \leftrightarrow E_x$ transition (Readout). Ideally, this results in the detection of 1 or more photons for the $m_s=0$ state, and no detected photons for $m_s=\pm1$. The resulting readout fidelities are asymmetric: $F_0 = 0.890(4)$ for $m_s=0$ (limited by the detection efficiency and number of cycles before a spin flip) and $F_1 = 0.988(2)$ for $m_s=\pm1$ (limited by background counts and unwanted excitations). Because uncontrolled spin flips in the excited state decohere nearby nuclear spins, we minimize the number of unnecessary optical excitations by using a weak readout pulse with a maximum duration of 114 $\mu$s ($\sim100$ excitations) and by switching off the laser within 2 $\mu$s ($\sim2$ excitations) once a photon is detected [3]. The resulting measurement is non-destructive: the probability that the spin prepared in $m_s=0$ is still in that state after a measurement with outcome $m_s=0$ is 0.992. In contrast, without dynamically stopping the laser the spin would be pumped almost completely to $m_s=\pm1$. For the final readout at the end of the experiment, which is allowed to be destructive, we use a stronger readout pulse of maximum duration 35 $\mu$s.       }
\label{Fig:OS} \vspace*{-0.2cm}
\end{figure} 

\begin{figure}[h!]
\begin{center}
\includegraphics[scale=0.70]{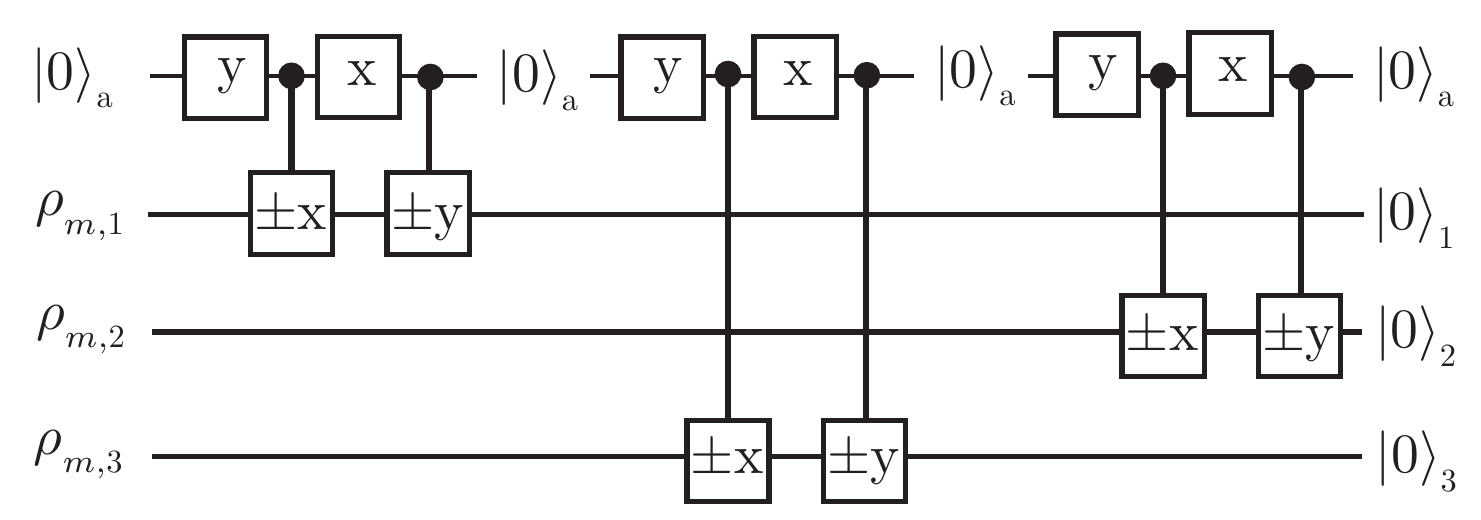}
\end{center}
\captionsetup{justification=raggedright}
\caption{\textbf{Qubit initialization.} All qubits naturally start in fully mixed states $\rho_m$. The ancilla is initialized in $\ket{0}_a$ and a reduced SWAP operation between the ancilla and the qubit is performed, deterministically initializing the qubit in $\ket{0}$. The ancilla is then reinitialized by a 300 $\mu$s laser pulse (Reset) and the process is repeated to initialize the other qubits.}
\label{Fig:init} \vspace*{-0.2cm}
\end{figure}

\begin{figure}[h!]
\begin{center}
\includegraphics[scale=0.60]{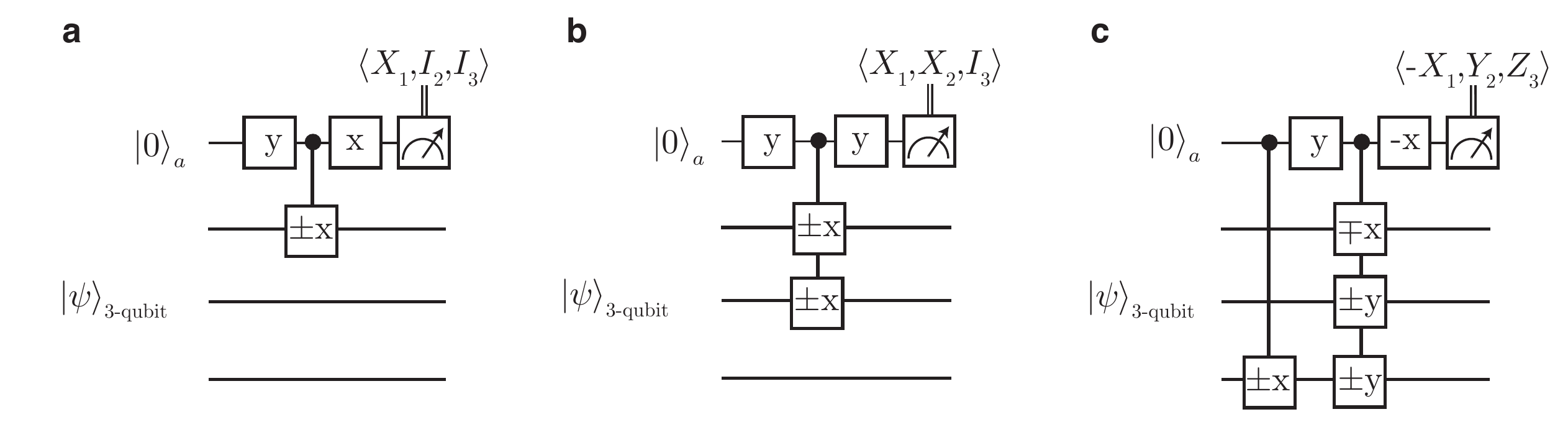}
\end{center}
\captionsetup{justification=raggedright}
\caption{\textbf{Tomography sequences for three-qubit states.} Examples of three-qubit expectation values that are measured by mapping the required correlation on the ancilla before reading it out. \textbf{a,} $\ev{X_1,I_2,I_3}$, 
\textbf{b,} $\ev{X_1,X_2,I_3}$, 
\textbf{c,} $\ev{-X_1,Y_2,Z_3}$. Note that the phase of the last $\pi/2$-pulse on the ancilla depends on the number of qubits read out (i.e. the number of operators that are not $I$). 
The examples given here for the measurement of one- two- and three-qubit expectation values can be translated to any of the 63 measurements in the full three-qubit state tomography.}
\label{Fig:tomo} \vspace*{-0.2cm}
\end{figure}

\begin{figure}[!]
\begin{center}
\includegraphics[scale=0.550]{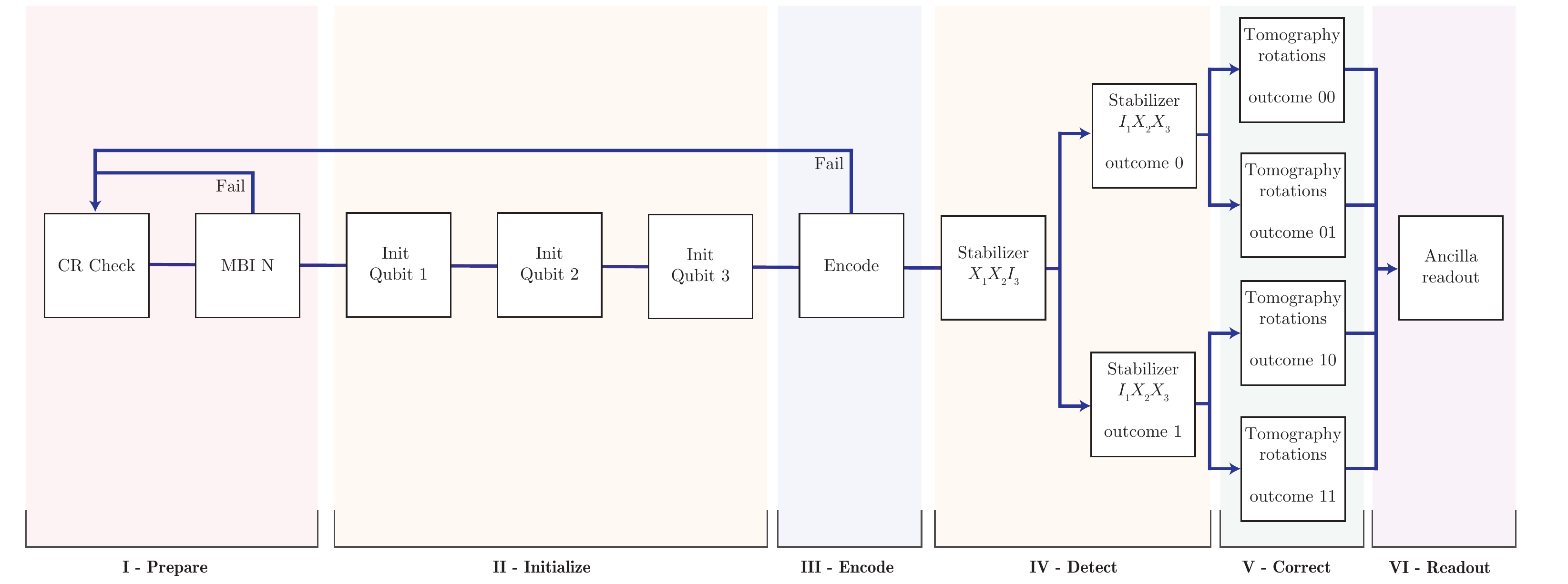}
\end{center}
\captionsetup{justification=raggedright}
\caption{\textbf{Experimental sequence and logic for the QEC experiments.}  Example for a single round of quantum error correction by stabilizer measurements (as in Fig. 3). The order of the sequence is controlled in real-time by an ADwin microprocessor. \textbf{I -} The NV center is prepared in its negative charge state and on resonance with the readout and reset lasers (Fig.~\ref{Fig:OS}) by turning on both lasers, counting the fluorescence photons and requiring a threshold to be passed (``CR Check''). The $^{14}N$ nuclear spin is initialized by measuring it and continuing only for outcome $m_I=-1$ (``MBI N'').      
\textbf{II -} The qubits are sequentially deterministically initialized following Fig.~\ref{Fig:init}.
\textbf{III -} The encoding of the logic state is a probabilistic process, as shown in Fig. 2a. When the wrong outcome is obtained the preparation of the experiment starts over.
\textbf{IV -} Errors are detected by two stabilizer measurements. Depending on the outcome ($-1$ or $+1$) of each measurement, the next sequence to execute is communicated to the waveform generator in real time. \textbf{V -} Depending on which of the 4 outcomes is obtained, a set of gates is performed to correct errors and to map the desired expectation value onto the ancilla (Fig.~\ref{Fig:tomo}). \textbf{VI -} Finally the ancilla is read out. Each outcome is taken into account without post processing or post selection. Note that for the experiment with three rounds of error correction (two rounds of stabilizer measurements QEC, Fig. 4b), the sequence branches in 16 paths instead.}
\label{Fig:flow} \vspace*{-0.2cm}
\end{figure}

\begin{figure}[h!]
\begin{center}
\includegraphics[scale=0.50]{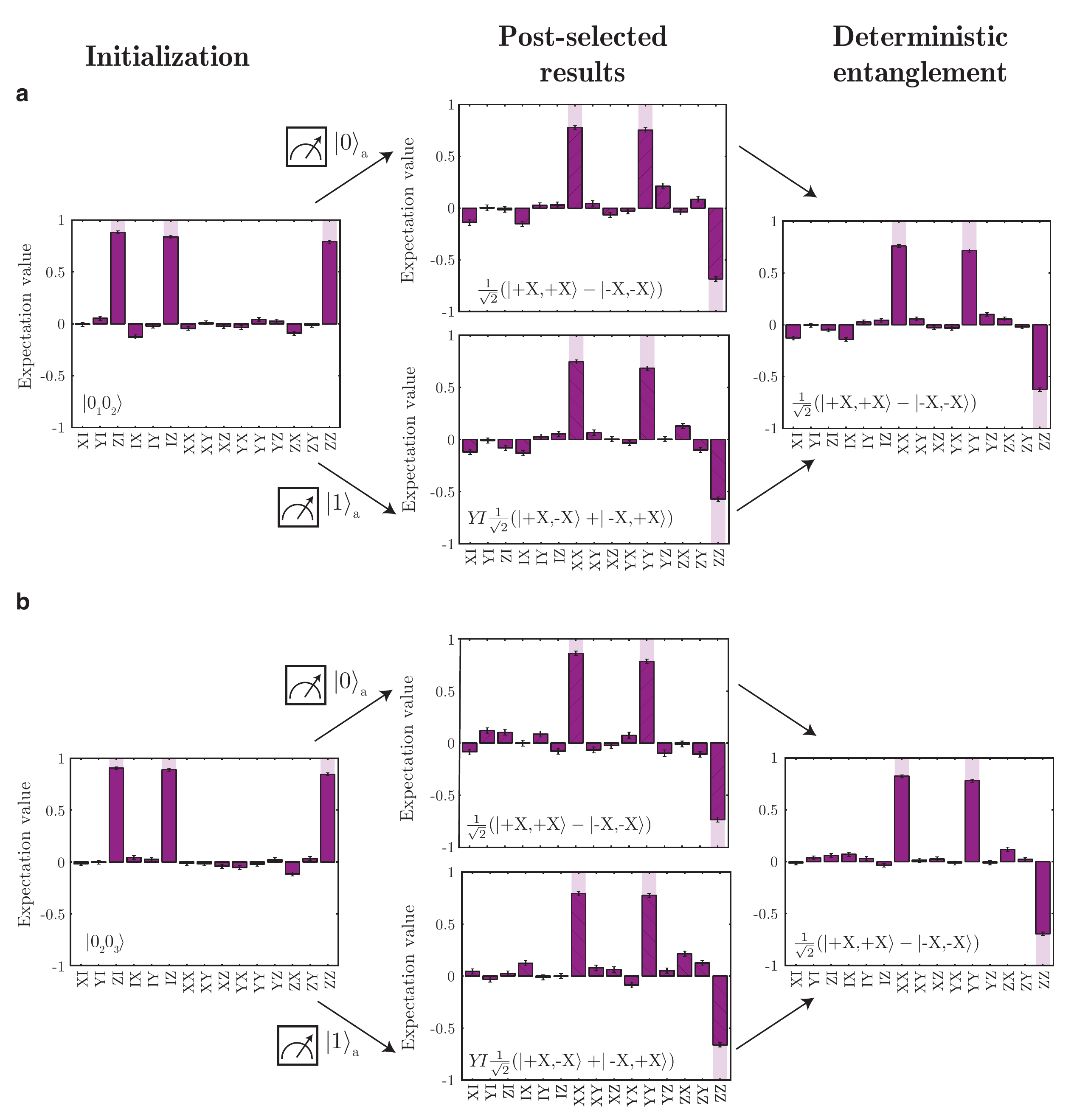}
\end{center}
\captionsetup{justification=raggedright}
\caption{\textbf{Deterministic entanglement by stabilizer measurements including post-selected results.} \textbf{a,} For qubits 1 and 2. \textbf{b,} For qubits 2 and 3. First the qubits are initialized following Fig.~\ref{Fig:init} in $\ket{00}$ with fidelity 0.878(6) for (a) and 0.910(6) for (b) (left column). Then a $XX$ measurement is performed (Fig. 1d). Depending on the measurement outcome feedback is applied, so that independent of the outcome the same two-qubit state is obtained, as can be seen by post-selecting on the two outcomes (middle column). The full result is a deterministically entangled state (right column). The fidelity with the desired two-qubit entangled state is 0.776(7) in (a) and 0.824(7) in (b).}
\label{Fig:MBE} \vspace*{-0.2cm}
\end{figure}

\begin{figure}[h!]
\begin{center}
\includegraphics[scale=0.80]{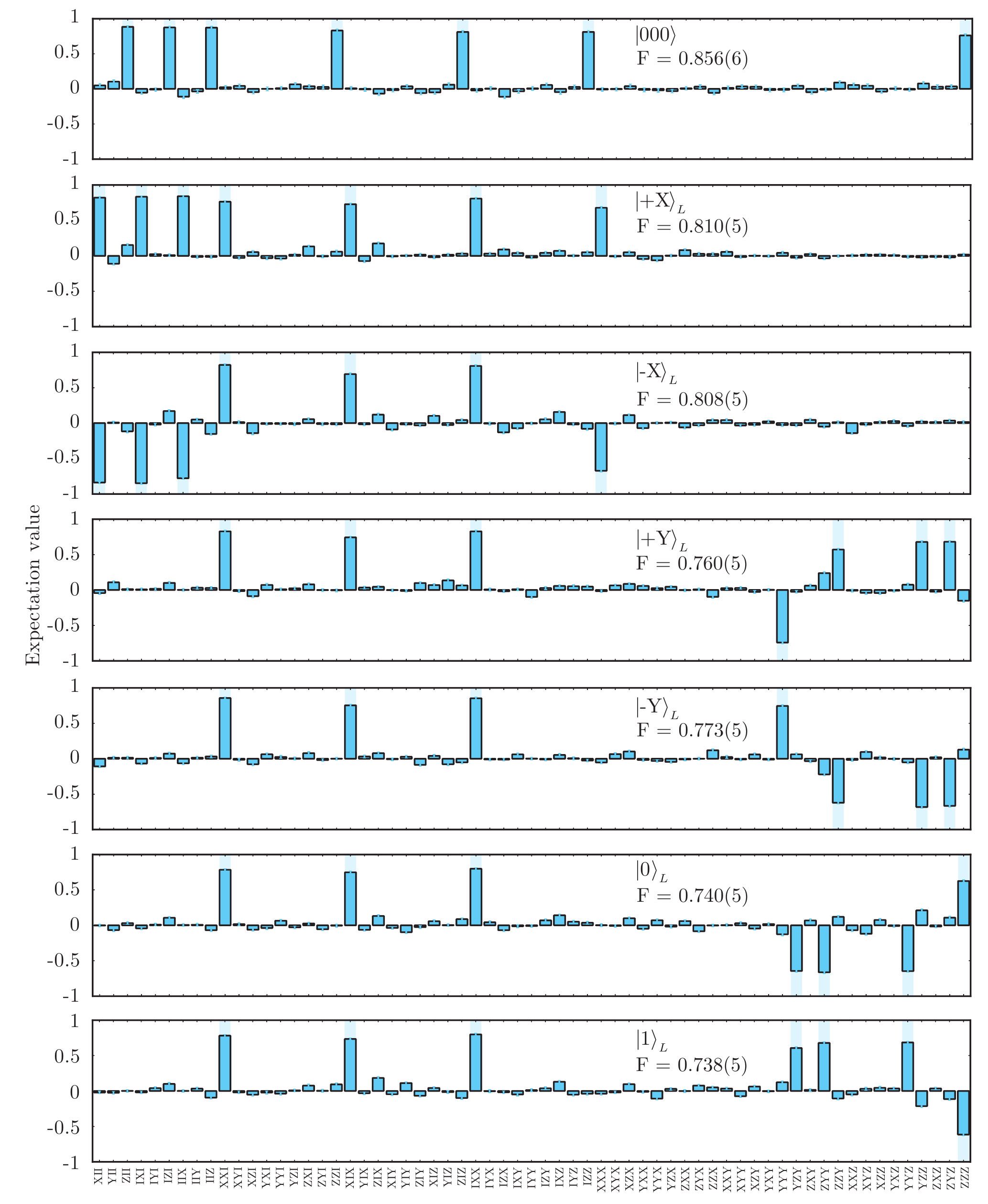}
\end{center}
\captionsetup{justification=raggedright}
\caption{\textbf{Three-qubit state tomography for $\ket{000}$ and the six logical states.} The three qubits are initialized as in Fig.~\ref{Fig:init} and encoded following Fig. 2a. The fidelities $F$ shown are the three-qubit state fidelities and the shaded bars indicate the ideal states. Ideally, the 6 encoded states are all eigenstates of the $XX$ stabilizers with eigenvalues $+1$, in agreement with the high values for $\ev{X_1,X_2,I_3}$,  $\ev{X_1,I_2,X_3}$ and  $\ev{I_1,X_2,X_3}$ for all states and an average fidelity with this code subspace of 0.839(3).
The logical qubit is encoded as $\alpha\ket{0}_L+\beta\ket{1}_L$, with $\ket{0}_L=\frac{1}{\sqrt{2}}(\kxy{+X,+X,+X}+\kxy{-X,-X,-X})$ and 
$\ket{1}_L=\frac{1}{\sqrt{2}}(\kxy{+X,+X,+X}-\kxy{-X,-X,-X})$. The logical state expectation values are given by:
 $\ev{X}_L=\ev{X_1,I_2,I_3}$,
 $\ev{Y}_L=\ev{Y_1,Z_2,Z_3}$,
$\ev{Z}_L=\ev{Z_1,Z_2,Z_3}$ 
or cyclic permutations. The logical qubit fidelities for the states are $\ket{+X}_L: 0.916(6)$, $\ket{-X}_L: 0.911(6)$, $\ket{+Y}_L: 0.822(7)$, $ \ket{-Y}_L: 0.828(7)$, $\ket{0}_L: 0.813(9)$ and $\ket{1}_L: 0.808(9)$.}
\label{Fig:ENC} \vspace*{-0.2cm}
\end{figure}




\begin{figure}[h!]
\begin{center}
\includegraphics[scale=0.40]{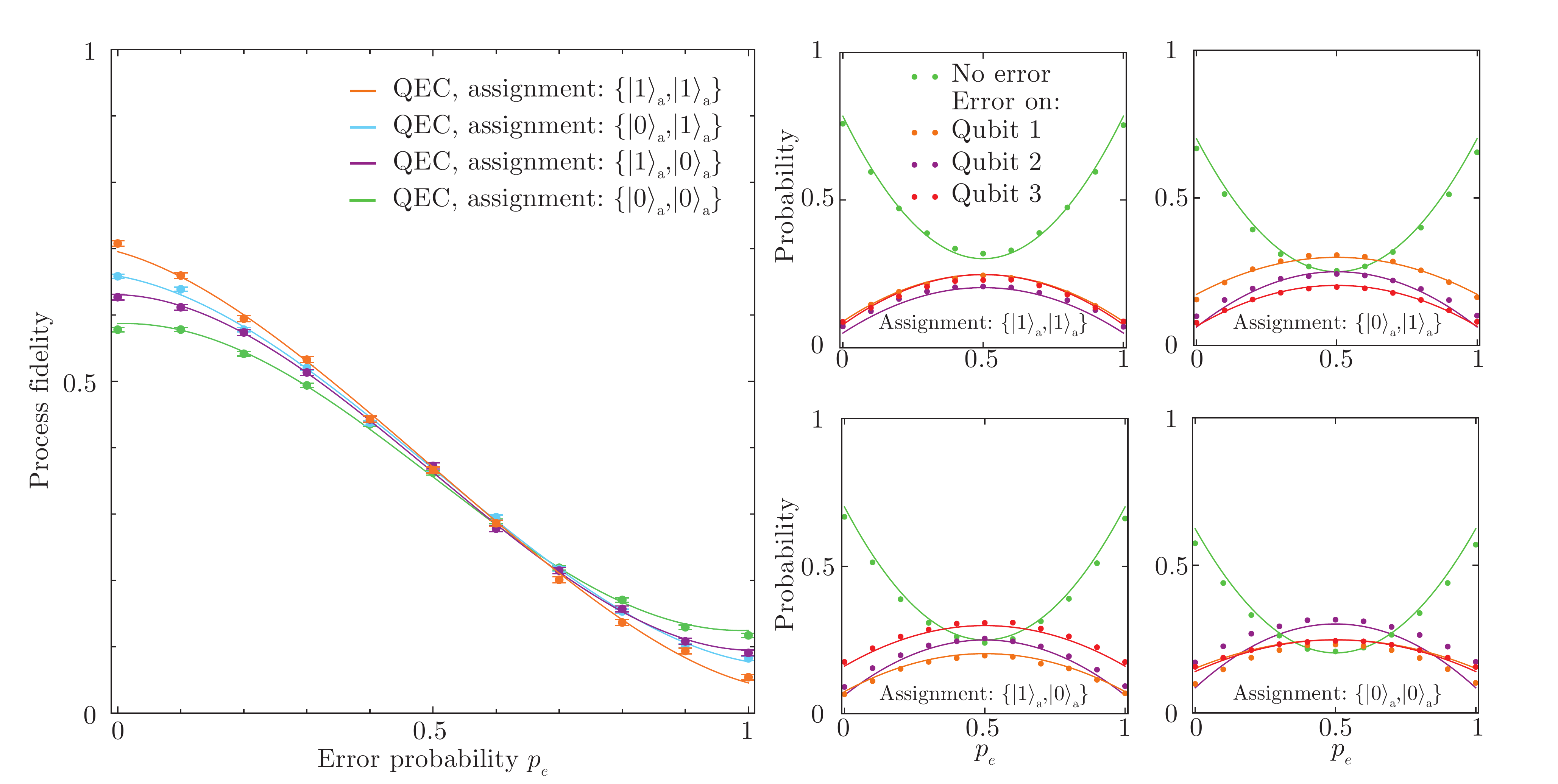}
\end{center}
\captionsetup{justification=raggedright}
\caption{\textbf{Process fidelity and error syndrome probabilities for different ancilla assignments.} Left: comparison of the process fidelities for the quantum error correction experiment in Fig. 3 for all four possible assignments of the ancilla states to the +1,+1 outcome of the stabilizer measurements. In Fig. 3b we average over these four curves. In Fig. 3c the optimal result is used (assignment $\{\ket{1}_a$, $\ket{1}_a\}$). Solid lines are fits to Eq. \ref{Eq:QEC} taking into account Eq. \ref{Eq:ED} and yield: 
$w = 0.8(1)$, 
$w = 0.71(7)$, 
$w= 0.95(7)$ and $w = 0.84(9)$ for the four assignments. Right: the probabilities for the error syndromes for each of the four ancilla state assignments. Solid lines are expected curves similar to Eqs. \ref{Eq:no_11}-\ref{Eq:3_11}, based on the estimated initial errors in the encoded states: $p^{(1)}_{\text{in}} = 0.091(2)$, $p^{(2)}_{\text{in}} = 0.064(2)$, $p^{(3)}_{\text{in}} = 0.077(2)$ obtained from Fig. \ref{Fig:ENC}. The theoretical probabilities are in good agreement with the experimental values (no free parameters). The probabilities are the normalized occurrences in 84000 samples for the assignments $\{\ket{0}_a,\ket{0}_a\}$ and $\{\ket{0}_a,\ket{1}_a\}$ and in 28000 samples for the assignments $\{\ket{1}_a,\ket{1}_a\}$ and $\{\ket{1}_a,\ket{0}_a\}$.}
\label{Fig:Syn} \vspace*{-0.2cm}
\end{figure}

\begin{figure}[h!]
\begin{center}
\includegraphics[scale=0.40]{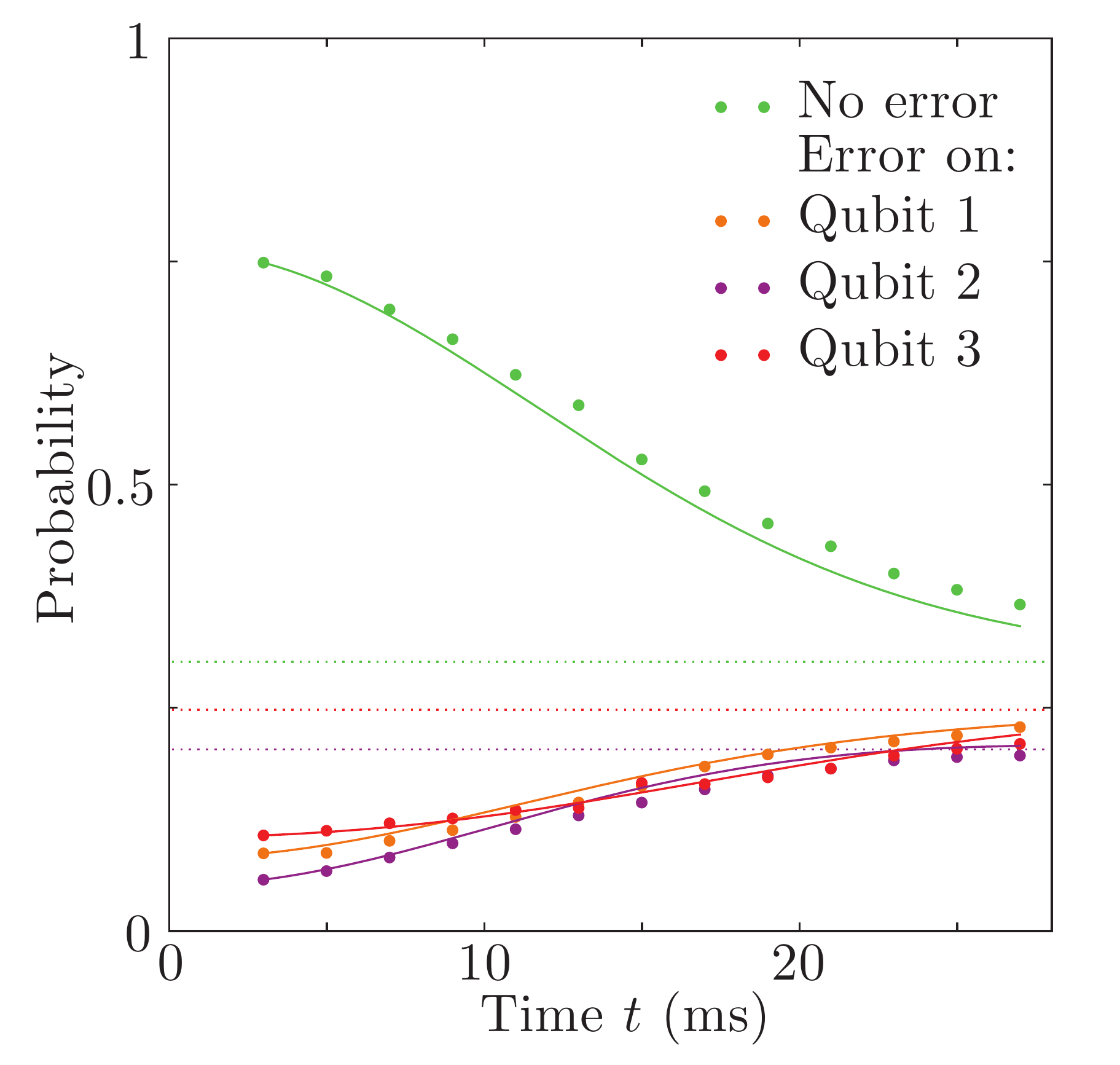}
\end{center}
\captionsetup{justification=raggedright}
\caption{\textbf{Error syndrome probabilities for naturally occurring errors.} Corresponding to Fig. 4d. Solid lines are theoretical predictions from the unique coherence times $T_2^*$ of the individual qubits and the initial error ($p_{\text{in}}$) determined from this data. As the stabilizer measurements are performed halfway the waiting time, the error probability for each qubit is: $p_e(\frac{t}{2}) = \frac{1}{2}(1-\text{Exp}[-(\frac{t}{2T_2^*})^2])$. Using Eqs.~\ref{Eq:no_11}-\ref{Eq:3_11} and the measured error outcome probabilities at the first datapoint ($t=2.99$ ms), we estimate the input errors at $t=0$ to be $p^{(1)}_{\text{in}} = 0.049(2)$, $p^{(2)}_{\text{in}} = 0.0804(4)$ and $p^{(3)}_{\text{in}} = 0.110(2)$. Dashed lines show the expected probabilities  for complete dephasing. The probabilities are based on the normalized occurrences in 12000 samples.
}
\label{Fig:E_ts} \vspace*{-0.2cm}
\end{figure}

\begin{figure}[h!]
\begin{center}
\includegraphics[scale=0.45]{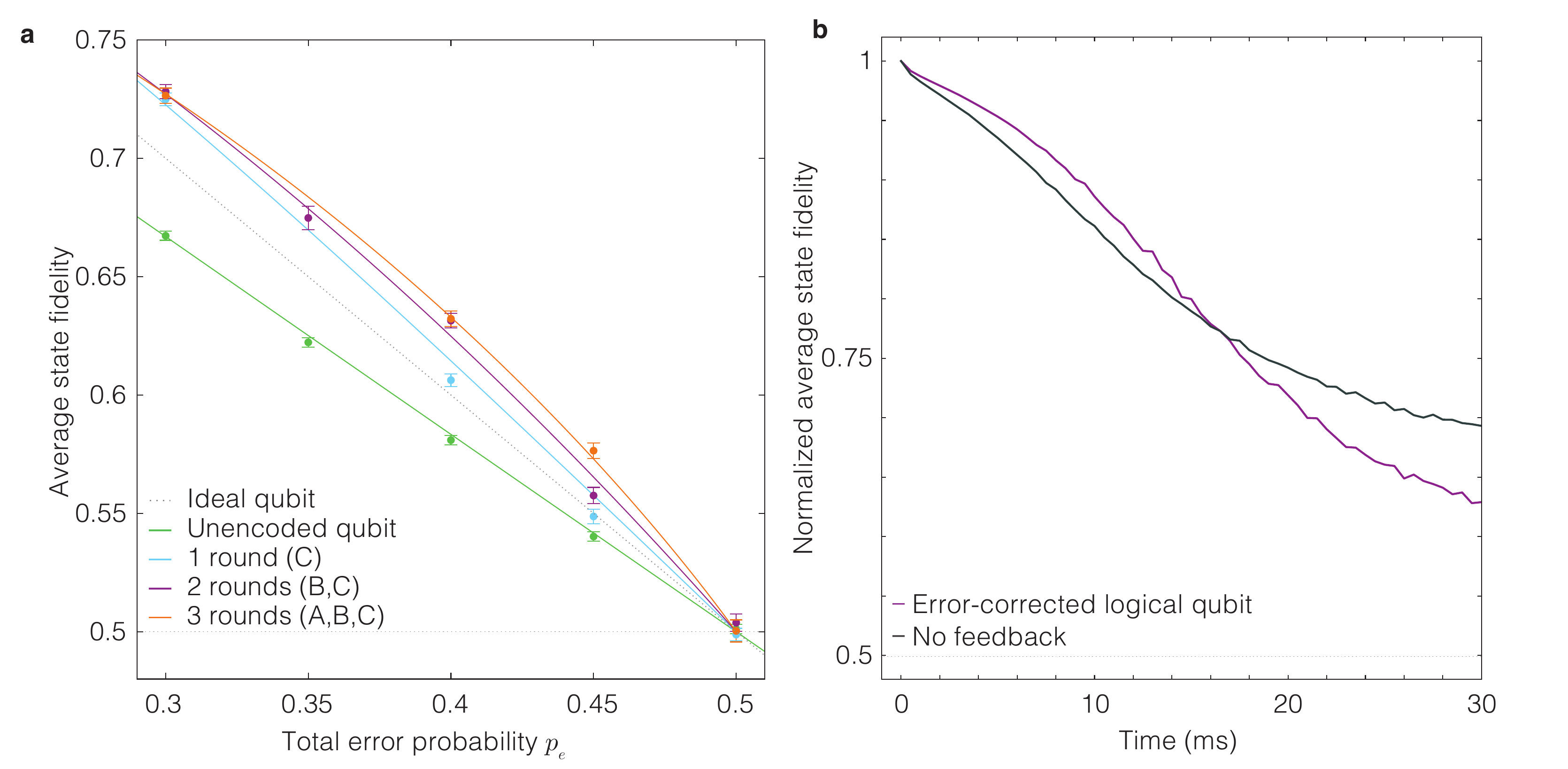}
\end{center}
\captionsetup{justification=raggedright}
\caption{\textbf{Zoom of data of Fig. 4b (a) and numerical simulation of Fig 4d (b).} \textbf{a,} A zoom-in of the area of the data in Fig. 4b in which additional rounds of error-correction are advantageous. \textbf{b,} Numerical Monte-Carlo simulations for the error correction experiment of Fig 4d. The initial state is $\ket{X}_L$. Each qubit then coherently evolves with a constant detuning randomly drawn from a Gaussian probability distribution with $\sigma = \sqrt{2}/T_2^*$. We take into account the measured longitudinal relaxation of each qubit, which approximately decays with $e^{-\sqrt{t/T_{1}}}$  for ancilla state $\ket{0}_a$ (See Tab.~\ref{Tab:Qubits}), the error-dependent readout fidelity (Eqs.~\ref{Eq:QEC}-\ref{Eq:p_dependence_of_A}), and a longitudinal relaxation of the ancilla with time-constant $300$~ms. Finally, detected errors are corrected (this last step is omitted for ``No feedback''). The simulation results qualitatively match all the main features of the observed dephasing curves (Fig.~4d). For short times, the stabilizer measurements suppress errors by stopping small errors from building up coherently and error correction further reduces the remaining errors. For long times, the stabilizer measurements halfway the sequence preferentially suppress coherent evolutions that would result in an error at the end of the sequence. As a result the fidelity at long times exceeds $0.5$ and decays only slowly. Moreover, applying error correction becomes detrimental: at the moment the stabilizer measurements are applied the state is essentially random and no useful information about errors is extracted so that applied corrections further dephase the final state. A complete quantitative comparison would require detailed modelling of the full evolution of the 4-qubit system during the gates, initialization, stabilizer measurements, and readout sequences as well as of the longitudinal decay at short times.}
\label{Fig10} \vspace*{-0.2cm}
\end{figure}



\begin{figure}[h!]
\begin{center}
\renewcommand{\figurename}{Supplementary Figure}
\includegraphics[scale=0.45]{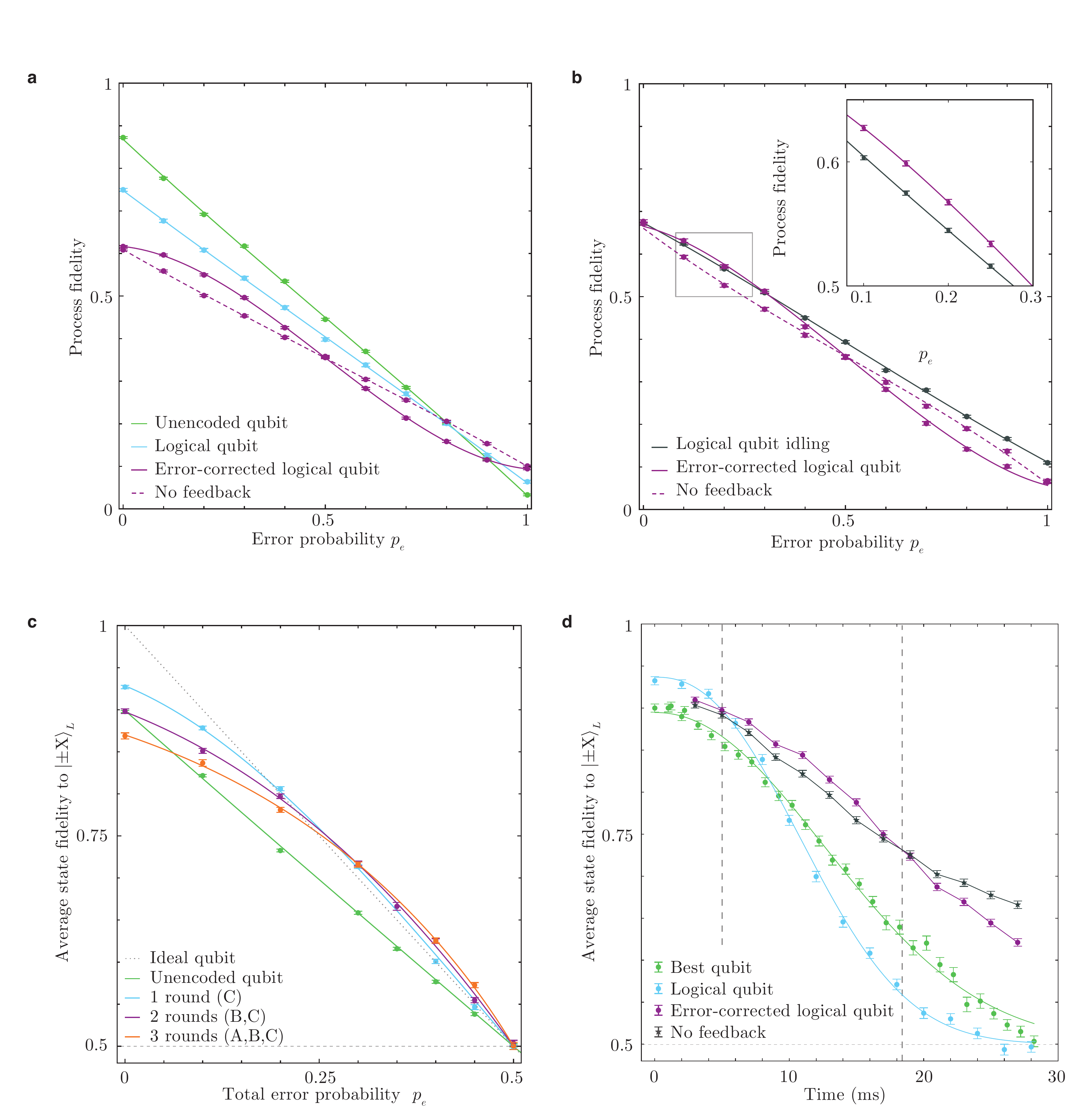}
\end{center}
\captionsetup{justification=raggedright}
\caption{\textbf{Data of Fig. 3b (a) 3c (b), 4b (c) and 4d (d), without correction for the final readout gates.} \textbf{a, b,} The fitted value for $w$ is indentical as in the main text (curve shapes are not influenced by the readout calibration). \textbf{c,} Corresponding fit values for the unencoded qubit: $w=-0.02(2)$, for 1 round: $w=0.56(6)$, for 2 rounds $w=0.64(4)$ and for three rounds $w=0.70(2)$. \textbf{d, } The logical qubit follows Eq.~\ref{Eq:decay} with $T=13.7(1)$ ms and $n=2.35(8)$.}
\label{Fig:fig3_uncorrected} \vspace*{-0.2cm}
\end{figure}

\FloatBarrier
\newpage
\section{Supplementary References}
\begin{itemize}

\item[[1]] Taminiau, T. H., Cramer, J., van der Sar, T., Dobrovitski, V. V. \& Hanson, R. Universal control and error correction in multi-qubit spin registers in diamond. \emph{Nature Nanotech.} \textbf{9}, 171-176 (2014).
\item[[2]] Robledo, L. \emph{et al}. High-fidelity projective read-out of a solid-state spin quantum register. \emph{Nature} \textbf{477}, 574-578 (2011).   
\item[[3]] Blok, M. S. \emph{et al}. Manipulating a qubit through the backaction of sequential partial measurements and real-time feedback.  \emph{Nature Phys.} \textbf{10}, 189-193 (2014).
\end{itemize}

\end{document}